\documentclass[aps,prb,reprint,superscriptaddress,nofootinbib,floatfix]{revtex4-2}
\usepackage{footnote}
\usepackage{blindtext} 
\usepackage{mathtools} 
\usepackage{graphicx}
\usepackage[export]{adjustbox}
\graphicspath{{./}{figures/}}
\usepackage{amsmath} 
\usepackage{amsfonts} 
\usepackage{amssymb}
\usepackage{amstext} 
\usepackage[english]{babel} 
\usepackage{helvet}
\usepackage{microtype} 
\usepackage[pdftex,hidelinks]{hyperref}
\hypersetup{
    colorlinks=true,
    linkcolor=blue,
    filecolor=blue,
    urlcolor=blue,
    citecolor = blue,
}
\usepackage{bbold} 
\usepackage{mathrsfs} 
\usepackage{makecell}

\newcommand{\rgeq}{\stackrel{\text{RG}}{=}}

\newcommand{\svdeq}{\stackrel{\text{svd}}{=}}

\newcommand{\textapproxmy}[1]{\stackrel{\text{#1}}{\approx}}
\newcommand{\texteq}[1]{\stackrel{\text{#1}}{=}}
\newcommand{\tTr}{\mathrm{tTr}}

\begin{document}
\title{Scaling dimensions from linearized tensor renormalization group transformations}

%Authors
\author{Xinliang Lyu} \email[]{lyu@issp.u-tokyo.ac.jp}
%\homepage[]{Your web page} \thanks{} \altaffiliation{}
\affiliation{Institute for Solid State Physics, The University of Tokyo,
Kashiwa, Chiba 277-8581, Japan} \author{RuQing G. Xu}
%\email[]{r-xu@g.ecc.u-tokyo.ac.jp}
\affiliation{Department of Physics, The University of Tokyo, Tokyo
    113-0033, Japan} \author{Naoki Kawashima}
    \email[]{kawashima@issp.u-tokyo.ac.jp} \affiliation{Institute for
        Solid State Physics, The University of Tokyo, Kashiwa, Chiba
    277-8581, Japan}

\date{\today}

\begin{abstract} 
    We show a way to perform the canonical renormalization group (RG) prescription in tensor space: write down the tensor RG equation, linearize it around a fixed-point tensor, and diagonalize the resulting linearized RG equation to obtain scaling dimensions. 
    The tensor RG methods have had a great success in producing accurate free energy compared with the conventional real-space RG schemes.
    However, the above-mentioned canonical procedure has not been implemented for general tensor-network-based RG schemes.
    We extend the success of the tensor methods further to extraction of scaling dimensions through the canonical RG prescription, without explicitly using the conformal field theory.
    This approach is benchmarked in the context of the Ising models in 1D and 2D.
    Based on a pure RG argument, the proposed method has potential applications to 3D systems, where the existing bread-and-butter method is inapplicable.
\end{abstract}

\maketitle

% body of paper here - Use proper section commands References should be
% done using the \cite, \ref, and \label commands
\section{Introduction\label{intro}}
The renormalization group (RG) is a powerful technique for studying physical systems where fluctuations in all scales of length are important~\cite{wilsonNobel}; the most famous example in statistical mechanics is critical phenomena.
The main idea behind the RG is to study how a physical system changes as we go from one length scale to another.
Conventional RG schemes, such as $\epsilon$-expansion~\cite{wilson1972} and block-spin methods~\cite{kadanoff1966,kadanoff1975,migdal,kadanoff1976,niemeijer1973}, aim at a map from the Hamiltonian of the short length scale to that of the longer one, such that the partition function is unchanged~\cite{nonlinearRG}.
The map is known as an RG equation. 
A well-behaved RG equation exhibits fixed points, each corresponding to a conformal field theory (CFT)~\cite{polchinski1988,nakayama2015}.
A critical system is described by a fixed point.
By linearizing the RG equation around the critical fixed point, universal properties like scaling dimensions of the critical system can be extracted.
This canonical RG prescription also provides a theoretical framework to understand universality in critical phenomena.
However, for a systematic study with high-precision, the Hamiltonian may not be the most efficient representation of the system.

Recently, ideas from quantum information have stimulated a novel type of RG methods in tensor space.
They are versatile numerical RG schemes whose approximations are controlled by an integer, $\chi$, called the \textit{bond dimension}.
The RG equation is a map from a tensor encapsulating the Boltzmann weights of local configurations at a short length scale to a new tensor at a longer one.
The first realization of this new paradigm is the tensor renormalization group (TRG)~\cite{trg}, followed by many variations~\cite{SRGa,SRGb,hotrg,atrg, triadtrg,morita2020global}.
These TRG-type techniques have excellent performance in calculations of free energy. 
For example, the higher-order tensor renormalization group (HOTRG)~\cite{hotrg} estimates the free energy of the 2D Ising model with error of order $10^{-7}$ within a few minutes in a desktop computer.
The estimation error decreases exponentially as $\chi$ increases, while the computational costs only grow polynomially.

With all of their success in calculations of free energy, however, the TRG-type techniques encounter obstacles in the canonical RG prescription. 
Early attempts~\cite{Berker2008,aoki2009,meurice2013,kadanoff2014} show that if the bond dimension $\chi$ of the TRG is larger than 8, the tensor will \textit{never} flow to the critical fixed point of the 2D Ising model; this imposes a very strong restriction on the bond dimension in the canonical RG prescription.
For $\chi = 2, 3, 4$, either using the TRG or the HOTRG, the estimated scaling dimension of the energy density operator has accuracy similar to the old potential moving tricks~\cite{Berker2008,aoki2009,meurice2013}, and that of the spin operator is more than a factor of 2 larger than the exact value~\cite{kadanoff2014}.

Fortunately, in recent ten years, people have developed many tricks to solve the problem of the unsatisfactory tensor RG flows.
In 2009, Gu and Wen~\cite{GuWen2009} was the first to deal with this problem. 
They followed Levin's suggestion~\cite{trg,LevinTalk} and focused on a toy model called corner double-line (CDL) tensors, which represent systems with only local correlations. 
They showed that the CDL tensors are fixed points of the TRG, indicating that the local correlations at the smaller length scales will be carried to the larger ones. 
A crude algorithm was proposed to filter out the CDL tensors and the problem of the tensor RG flows was partially solved, followed up by an improved algorithm in 2017~\cite{looptnr}. 
From 2015 to 2017, several similar methods were proposed~\cite{tnr,tnralgo,tnrplus}. 
All of these advanced TRG-type techniques successfully produced critical fixed-point tensors. 

With a critical fixed-point tensor in hand, Gu and Wen~\cite{GuWen2009} pointed out that the scaling dimensions can be extracted by diagonalizing a transfer matrix constructed from the fixed-point tensor according to a well-known 2D CFT theorem~\cite{cardy1986}. 
Gu and Wen's proposal gradually becomes the bread-and-butter method for TRG-type techniques to extract scaling dimensions of 2D systems. 
Later, Evenbly and Vidal used the tensor network renormalization (TNR)~\cite{tnr,tnralgo} to implement local scale transformation that maps a plane to a cylinder~\cite{EvenblyDilatationOp}; the spectrum of eigenvalues of a transfer matrix on the cylinder gives scaling dimensions.
These two methods have been applied to extract scaling dimensions from then on, while the canonical RG prescription in tensor space has never followed up.
%
% Figure: research background
\begin{figure}[tb]
    \includegraphics[width=1.0\columnwidth,valign=c]{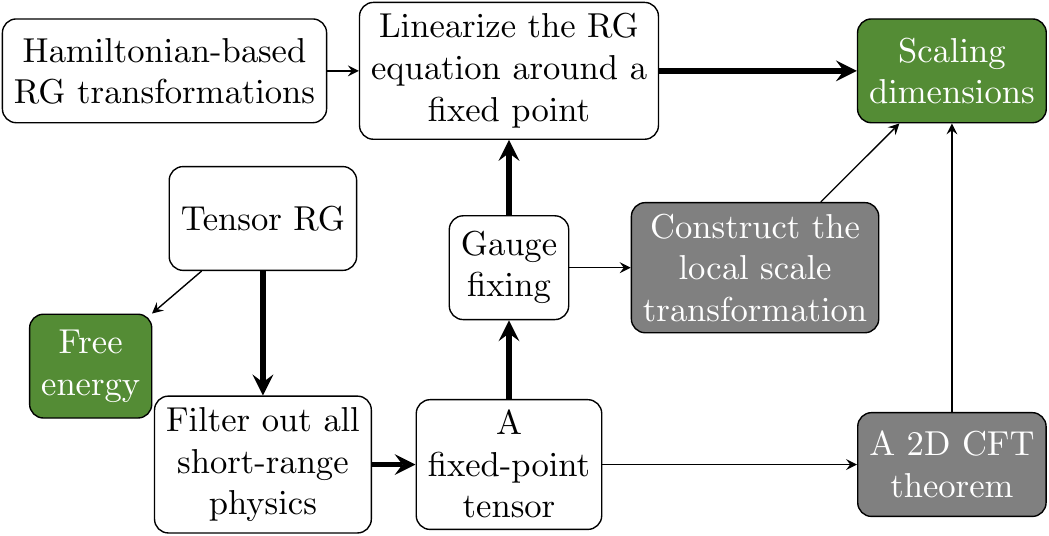}
    \caption{\label{fig:researchBG}
        Different ways to extract scaling dimensions using tensor RG methods.
        The proposed method in this paper corresponds to the path indicated by the thick arrows.
    }
\end{figure}
% Figure end

\begin{figure}[tb]
    \includegraphics[scale=0.9,valign=c]{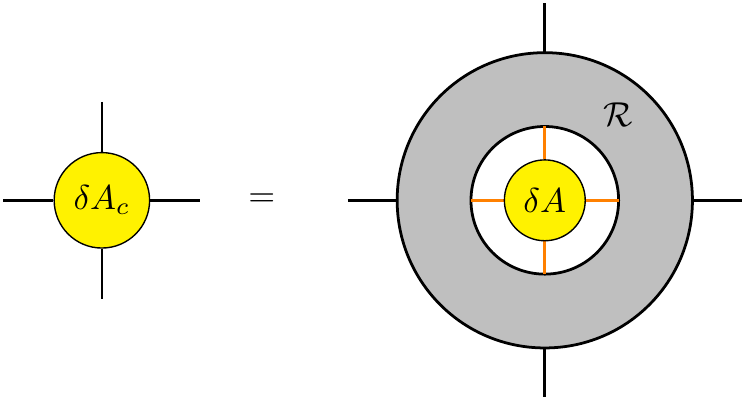}
    \caption{\label{fig:linearedRGschem}
        Schematic diagram of the linearized tensor RG equation $\mathcal{R}$. 
        It is a linear map from $\delta A$ to $\delta A_c$ and is determined by the fixed-point tensor $A^*$ and various related tensors shown explicitly in Eq.~\eqref{eq:respMatGiltHOTRG}.
        The computation costs of this linearized RG equation are $O(\chi^7)$, the same as those of the tensor coarse graining, since the inner structure of $\mathcal{R}$ resembles the tensor network in the coarse graining step in Eq.~\eqref{def:RGeqGiltHOTRG}.
    }
\end{figure}

In this paper, we provide the missing piece of carrying out, in the general prescription, the RG in tensor space at a general bond dimension (see Fig.~\ref{fig:researchBG}).
After laying down the general framework for the canonical RG prescription for the tensor RG methods in Sec.~\ref{RGprescrip}, we point out two technical obstacles, local correlations and gauge redundancy, in Sec.~\ref{sec:obstacles}.
The higher-order tensor renormalization group (HOTRG)~\cite{hotrg} is combined with a recently-developed technique, graph-independent local truncation (GILT)~\cite{gilts} in Sec.~\ref{sec:gilthotrg}, to generate correct tensor RG flows that will go to a critical fixed point at a general bond dimension. 
In Sec.~\ref{sec:gaugefix}, we show that most gauge redundancy in the tensor description is automatically fixed in the proposed HOTRG-like scheme, leaving only tractable sign ambiguities.
The linearized RG equation of this HOTRG-like scheme is easy to implement and has a simple pictorial representation\footnote{It should be noted that the linearized RG transformation looks similar to the local scale transformation of Ref.~\cite{EvenblyDilatationOp}.
        Since the canonical RG prescription here is applicable to any proper tensor RG methods, it is more general than the TNR local scale transformation.
    }
(see Fig.~\ref{fig:linearedRGschem} and Eq.~\eqref{eq:respMatGiltHOTRG}); in practice, it can be generated by automatic differentiation~\cite{xiang2019adtrg} once the tensor RG equation is implemented.
The scaling dimensions can be extracted from this linearized RG equation.
In Sec.~\ref{benchmark}, the canonical RG prescription in tensor space is benchmarked with the 1D and 2D classical Ising models.
We conclude in Sec.~\ref{conclusion}.
%
% Figure: schematic figure of linearized RG
% Figure end

The proposed method fully exploits the RG interpretation of the TRG-type techniques and offers a better understanding of the nature of these tensor techniques as real-space RG transformations. 
Since the HOTRG can be viewed as a modern extension of the Migdal-Kadanoff RG~\cite{migdal,kadanoff1976,meurice2013}, the canonical RG prescription based on the proposed HOTRG-like scheme develops the old Migdal-Kadanoff idea in tensor network language and makes it systematically improvable.
Most notably, the method might be relevant in 3D, where Gu and Wen's method is inapplicable and Evenbly and Vidal's local-scaling-transformation idea is nontrivial to implement.
%%%

\section{Renormalization group in tensor network language\label{sec:RGtensorSpace}} 
\begin{figure}[tb]
    \includegraphics[scale=0.8]{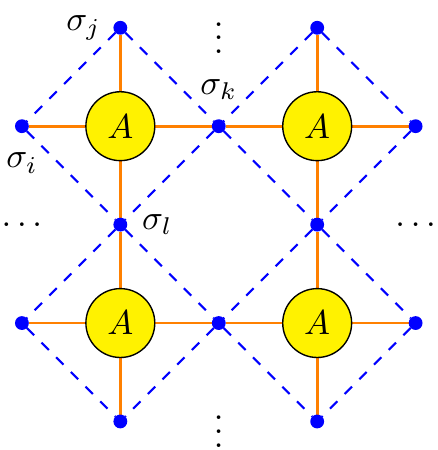}
    \caption{\label{fig:spin2tensor}
        Representation of the local Boltzmann weight in terms of tensors.
        The dots are where the spin variables locate.
        They form a square lattice slanted by $45^\circ$.
        The larger circles are tensors $A$ encoding the Boltzmann weight of the configurations of the four surrounding spin variables.
    The square lattice formed by $N$ copies of $A$ is the tensor network representation of the partition function in Eq.~\eqref{eq:ZbeforeRG}.
}
\end{figure}

TRG-type methods start with the fact that partition functions of all classical statistical models can be rewritten as tensor network models~\cite{trg}.
Take the square lattice 2D Ising model as a concrete example. The partition function is
\begin{align}\label{eq:2DIsingZ}
    Z =
\sum_{\{\sigma(\mathbf{r})\}}e^{K\sum_{\langle i,j \rangle}\sigma_i \sigma_j},
\end{align}
where $\sigma_i$ is the shorthand for the spin variable $\sigma(\mathbf{r}_i)$ located at lattice point $\mathbf{r}_i$ and can take values $\pm 1$, and $K = J / k_B T$.
In this paper, we measure temperature in units of $J / k_B $ so it becomes a dimensionless number.
The partition function in Eq.~\eqref{eq:2DIsingZ} can be rewritten as a tensor network by defining a tensor 
\begin{align}\label{def:tensorA}
    A_{\sigma_i \sigma_j \sigma_k \sigma_l}
\equiv e^{K(\sigma_i\sigma_j + \sigma_j\sigma_k + \sigma_k\sigma_l +
\sigma_l\sigma_i)}
= \includegraphics[scale=0.8,valign=c]{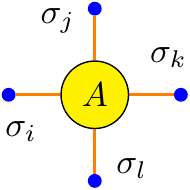}\text{ }.
\end{align}
Each index of this tensor can take two values $\pm 1$ and we say the bond dimension of a leg of this tensor is $\chi = 2$.
It is now possible to rewrite the partition function of the 2D Ising model in Eq.~\eqref{eq:2DIsingZ} as the tensor product of $N$ copies of $A$, with all their indices summed over (Fig.~\ref{fig:spin2tensor})
\begin{align}\label{eq:ZbeforeRG}
    Z = \sum_{\{ \sigma(\mathbf{r}) \}}
    \bigotimes^{N}_{x=1}A_{\sigma_{i(x)} \sigma_{j(x)} \sigma_{k(x)} \sigma_{l(x)}}
    \equiv \tTr\left(\bigotimes_{x=1}^{N}A\right),
\end{align}
where the last equal sign defines the $\tTr$ symbol.
%
% Figure: mapping from spins on lattice to a tensor network
% Figure end

The coarse graining of the tensor network resembles the conventional block-spin methods.
We replace a patch of, say, four copies of the original tensor $A$ with one coarse-grained tensor $A_c$, such that the partition function is approximately described by a coarser tensor network made of $N/4$ copies of $A_c$
\begin{align}\label{eq:ZafterRG}
    Z \approx \tTr\left(\bigotimes_{x=1}^{N/4}A_c\right).
\end{align}
The specific procedure for obtaining $A_c$ from $A$ will be discussed later in Sec.~\ref{sec:gilthotrg}. The map, 
\begin{align}\label{def:tensorRGeq}
    A \xrightarrow{\text{RG}} A_c \text{ or } A_c = \mathcal{T}(A), 
\end{align}
is the tensor RG equation if the gauge redundancy of the tensor is properly fixed (see Sec.~\ref{sec:obstacles} and Sec.~\ref{sec:gaugefix} for detailed discussions about gauge redundancy in tensor network language and how it is fixed).

\subsection{General framework\label{RGprescrip}}
We first define the \textit{canonical} RG prescription in tensor space.
To this end, it is helpful to start with a review of the old approach in Hamiltonian space (we follow the detailed review~\cite{kadanoff2014} and textbook~\cite{cardy_1996} closely).

It will be convenient to explain in terms of a specific physical system, a classical system with spin variables $\sigma \in \{+1, -1\}$ on a lattice, with general short-ranged interactions.
The Hamiltonian (or energy) of the system can be parameterized by a set of coupling constants $\mathbf{K} = \{K_j\}$, each of which couples to a possible short-ranged interaction term $s_j(\mathbf{r})$,
\begin{align}\label{eq:generalspinHam}
    \mathcal{H} = \sum_{\mathbf{r}} \sum_{j} K_j s_j(\mathbf{r}).
\end{align}
For example, if $K_1$ is the magnetic field, $s_1(\mathbf{r}) = \sigma(\mathbf{r})$ is the spin variable at lattice point $\mathbf{r}$; if $K_2$ is the nearest neighbor interaction along $x$ direction, $s_2(\mathbf{r}) = \sigma(\mathbf{r})\sigma(\mathbf{r} + a\hat{\mathbf{e}}_x)$, where $\hat{\mathbf{e}}_x$ is the unit vector along $x$ direction and $a$ is the lattice constant. 
A conventional RG transformation maps the old Hamiltonian $\mathcal{H}$ to a new one $\mathcal{H}'$ \textit{with the same form as} Eq.~\eqref{eq:generalspinHam} but characterized by a set of new coupling constants $\mathbf{K}' = \{ K_j'\}$. 
The map from the old Hamiltonian to the new one $\mathcal{H} \xrightarrow{\text{RG}} \mathcal{H}'$ is then parametrized explicitly as the transformation from the old coupling constants to the new ones,
\begin{align}\label{eq:oldRGK}
    \mathbf{K}' = \mathcal{T}^{\text{old}}\left(\mathbf{K}\right).
\end{align}
We require that the RG transformation should preserve the partition function of the system and should exhibit a fixed-point Hamiltonian $\mathcal{H}^{*}$ parameterized by coupling constants $\mathbf{K}^{*}$, such that $\mathbf{K}^{*}$ remains unchanged under the RG transformation,
\begin{align}\label{eq:oldRGKstar}
    \mathbf{K}^{*} =
    \mathcal{T}^{\text{old}}\left(\mathbf{K}^{*}\right).
\end{align}

The linearized RG equation around $\mathbf{K}^{*}$ is defined in the following way.
We perturb the coupling constants around the fixed point $\mathbf{K} = \mathbf{K}^{*} + \delta \mathbf{K}$ and perform the RG transformation defined in Eq.~\eqref{eq:oldRGK}. 
The new coupling constants $\mathbf{K}'$ after the RG transformation should be close to $\mathbf{K}^{*}$ by continuity, so $\mathbf{K}' = \mathbf{K}^{*} + \delta \mathbf{K}'$.
The linearized RG equation around $\mathbf{K}^{*}$ is a matrix $\mathcal{R}^{\text{old}}$ telling us how $\delta \mathbf{K}'$ is related to $\delta \mathbf{K}$,
\begin{align}\label{eq:respMat}
    \delta K_i' = \sum_j\mathcal{R}^{\text{old}}_{ij} \delta K_j.
\end{align}
The matrix $\mathcal{R}^{\text{old}}$ has right and left eigenvectors $\{\psi^{\alpha}\}, \{\phi^{\alpha}\}$ with the same set of eigenvalues $\{\lambda^{\alpha}\}$,
\begin{align}\label{eq:eigsofRespM}
    \sum_j \mathcal{R}^{old}_{ij} \psi^{\alpha}_j = \lambda^{\alpha}
    \psi^{\alpha}_i \text{ and } \sum_i \phi^{\alpha}_i
    \mathcal{R}^{old}_{ij} = \lambda^{\alpha} \phi^{\alpha}_j.
\end{align}
The linear combinations of $\delta K_i$ according to the components of the left eigenvector $\phi^{\alpha}$ are known as scaling fields
\begin{align}\label{def:scalingfields}
    h^{\alpha} = \sum_i \phi^{\alpha}_i \delta K_i,
\end{align}
while the linear combinations of interaction terms $s_j(\mathbf{r})$ according to the components of the right eigenvector $\psi^{\alpha}$ are known as scaling operators
\begin{align}\label{def:scalingOpt}
    o^{\alpha}(\mathbf{r}) = \sum_j s_j(\mathbf{r}) \psi^{\alpha}_j.
\end{align}
Under the RG transformation with rescaling factor $b$ for a system in dimension $d$, the scaling fields and the scaling operators transform in a simpler way with 
\begin{align}\label{eq:transfho}
    \left(h^{\alpha} \right)' = b^{d - x_{\alpha}} h^{\alpha} \text{ and }
    \left(o^{\alpha}\right)' = b^{x_{\alpha}} o^{\alpha},
\end{align}
where $x_{\alpha}$ are the scaling dimensions of the scaling operators $o^{\alpha}(\mathbf{r})$. 
Equations~\eqref{eq:respMat} to~\eqref{def:scalingfields} and~\eqref{eq:transfho} give the relation between the scaling dimensions $\{x_{\alpha}\}$ and the eigenvalues $\{\lambda^{\alpha}\}$ of the linearized RG equation, 
\begin{align}\label{eq:lambda2x}
    b^{d-x_{\alpha}} = \lambda^{\alpha}.
\end{align}

Next, we move on to the tensor approach of the canonical RG prescription.
In the tensor RG approach, we skip the Hamiltonian description of the system. 
Instead, we use a tensor network made of copies of tensor $A$ to represent the partition function $Z$ of the system. 
The tensor RG equation is a map from the tensor $A$ to the coarser tensor $A_c$, as is shown in Eq.~\eqref{def:tensorRGeq}.
We claim that the components of the tensor $A$ can be thought of as some proxies of the coupling constants $\mathbf{K}$ (this claim was hinted in Ref.~\cite{GuWen2009}).

To see why this claim is reasonable, note that we can map the partition function of the system with Hamiltonian in Eq.~\eqref{eq:generalspinHam} to a tensor network using the method introduced in Ref.~\cite{trg}. 
Each component of the initial tensor $A$ is the Boltzmann weight of a given local configuration and depends on the coupling constants,
\begin{align}\label{eq:K2A}
    A_{(i)} = f_{(i)}\left(\mathbf{K}\right),
\end{align}
where we group all legs of $A$ to form a single index, $A_{(i)}\equiv A_{i_1 i_2 i_3 i_4}$. 
After coarse graining, the components of $A_c$ are still functions of $\mathbf{K}$ but
with different functional forms,
\begin{align}\label{eq:tensorEleRG}
    \left(A_c\right)_{(i)} =
\left(f_c\right)_{(i)}\left(\mathbf{K}\right).
\end{align}
Now, we require that each component of the coarser tensor $A_c$ should have the same functional form as that of $A$, but with different coupling constants $\mathbf{K}'$,
\begin{align}\label{eq:tensorK2Kp}
    f_{(i)}\left(\mathbf{K}'\right) =
    \left(f_c\right)_{(i)}\left(\mathbf{K}\right), \forall (i).
\end{align}
In the old Hamiltonian approach, we need to solve Eq.~\eqref{eq:tensorK2Kp} for $\mathbf{K}'$ in terms of $\mathbf{K}$, which defines the RG equation from the old $\mathbf{K}$ to the new $\mathbf{K}'$. 
However, in the tensor approach, it is enough to know the existence of such $\mathbf{K}'$. Combine Eq.~\eqref{eq:tensorEleRG} and Eq.~\eqref{eq:tensorK2Kp}, we have
\begin{align}\label{eq:Kp2A}
    \left(A_c\right)_{(i)} = f_{(i)}\left(\mathbf{K}'\right).
\end{align}
At the fixed point, $\mathbf{K} = \mathbf{K}' = \mathbf{K}^*$, equations~\eqref{eq:K2A} and \eqref{eq:Kp2A} give
\begin{align}\label{eq:tensorRGAstar}
    A^* \xrightarrow{\text{RG}} A^* \text{ or }A^* = \mathcal{T}\left(A^* \right).
\end{align}
Take the total derivative of tensors $A$ and $A_c$ in Eqs.~\eqref{eq:K2A} and \eqref{eq:Kp2A} and set $\mathbf{K} = \mathbf{K}' = \mathbf{K}^*$, 
\begin{align}
    \delta A_{(i)} = \sum_n \left(\partial_{(n)}
    f_{(i)}\right)\Bigr|_{\mathbf{K} = \mathbf{K}^*} \delta K_n,
    \label{eq:deltaK2deltaA} \\
    \left(\delta A_c \right)_{(i)} = \sum_n \left(\partial_{(n)}
    f_{(i)}\right)\Bigr|_{\mathbf{K}' = \mathbf{K}^*} \delta
    K_n'.\label{eq:deltaK2deltaAp}
\end{align}
Equations~\eqref{eq:deltaK2deltaA} and \eqref{eq:deltaK2deltaAp} give the transformation law between the coupling-constant description and tensor description of the canonical RG prescription, with $\partial_{(n)}f_{(i)}$ evaluated at $\mathbf{K}^*$ being the change of basis matrix.
Under this transformation, the linearized RG equation in Eq.~\eqref{eq:respMat} becomes
\begin{align}\label{eq:respMatTen}
    \left(\delta A_c\right)_{(i)} = \sum_j
    \mathcal{R}_{(i)(j)} \delta A_{(j)},
\end{align}
which defines the linearized RG equation in tensor space.
Since Eqs.~\eqref{eq:respMat} and \eqref{eq:respMatTen} are the same linear transformation in two different representations, we can equally well diagonalize the matrix $\mathcal{R}_{(i)(j)}$ and find scaling dimensions according to Eq.~\eqref{eq:lambda2x}. 

In Sec.~\ref{benchmark:1DIsing}, we will use the 1D Ising model as a
concrete example to demonstrate the general argument above.

\subsection{Technical obstacles\label{sec:obstacles}}
There are two major obstacles for the canonical RG prescription in tensor space: the problem of local correlations and the gauge redundancy in tensor network language. 
They prevent us from obtaining a fixed-point tensor satisfying Eq.~\eqref{eq:tensorRGAstar}.

Levin and Nave anticipated the problem of local correlations when looking for fixed points of the RG equation of the TRG~\cite{LevinTalk}.
One of the earliest numerical evidence for the peculiar tensor RG flows of the 2D Ising model was provided by Hinczewski and Berker~\cite{Berker2008}. 
Their results indicate that the TRG-type techniques have difficulty in integrating out all the local correlations at short distances, so physics at the original lattice scale is carried all the way to the physics at larger ones. 
This shortcoming of the TRG-type techniques makes identification of both non-critical and critical fixed-point tensors very difficult.

\begin{figure}[t]
    \includegraphics[scale=0.9,valign=c]{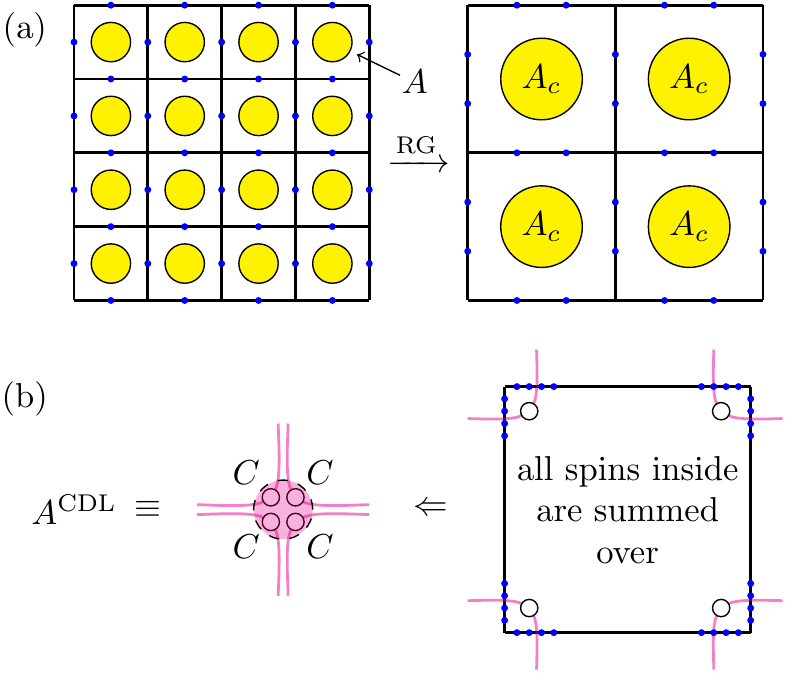}
    \caption{\label{fig:rgschem}
        The origin of the problem of local correlations.
        (a) The block-tensor transformation $A\rightarrow A_c$. 
        The spins shared by two tensors $A$ are summed over according to Eq.~\eqref{eq:exactBlock}. 
        The squares are larger after the decimation.
        (b) The origin of the
        CDL tensors. When the black square becomes large enough, the
        spins on one edge are far away from those on another, except for
        the spins around the four corners. The correlations among the
corner spins give rise to the CDL tensors, containing physics at the
lattice scale.
} 
\end{figure}

To understand how the problem of local correlations at the lattice scale arises in the TRG-type techniques, let us examine the physical picture of the tensor RG transformation.
We focus on a concrete example of a tensor network made of $4 \times 4 = 16$ copies of tensor $A$ shown in Fig.~\ref{fig:spin2tensor} with periodic boundary condition.
The general picture of a tensor RG transformation is similar to the conventional block-spin methods.
For example, we block a square of four tensors by contracting legs between them and group every two legs in the same side. Call the new tensor $A_c$,
\begin{align}\label{eq:ZunderRG} 
    Z_{4 \times 4} = 
    \includegraphics[scale=1.0,valign=c]{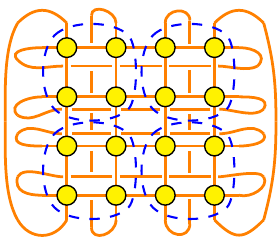}
    \rgeq 
    \includegraphics[scale=0.8,valign=c]{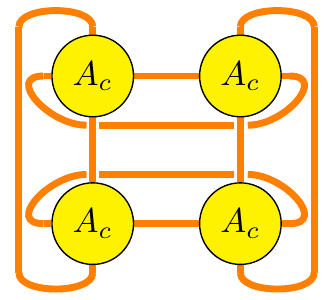},
\end{align}
where 
\begin{align}\label{eq:exactBlock}
    \includegraphics[scale=0.8,valign=c]{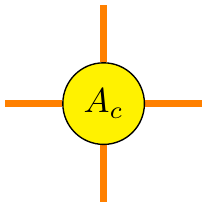}
    \equiv 
    \includegraphics[scale=0.8,valign=c]{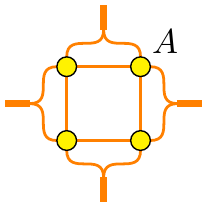}.
\end{align}
It is enlightening to put the original spin variables back into the tensor network to get a more physical picture of what is happening under such a block-tensor RG transformation. 
We refrain from drawing legs of $A$ and the dashed lines of the spin lattice in Fig.~\ref{fig:spin2tensor}, and surround copies of $A$ with squares on whose sides the spin variables sit. 
The big picture for the block-tensor transformation in Eq.~\eqref{eq:ZunderRG} is shown schematically in Fig.~\ref{fig:rgschem}(a).
%
% RG directly in both tensor and original spin decimation language
The process is similar to the decimation in the conventional approaches.
After the spin variables shared by every two $A$ tensors forming the same $A_c$ are summed over, we are left with four bigger squares, with two spin variables sitting on each side of each square.
When the squares become large enough as the block-tensor transformation goes on, we expect that, roughly speaking, the spin variables on different edges are far away from each other and thus uncorrelated. 
The only exception is for the spin variables around the four corners.
We can use a matrix $C$ in Fig.~\ref{fig:rgschem}(b) to capture the correlations around the corners; the matrix $C$ must contain physics at the scale of the original lattice constant. 
Since the spin variables around different corners are far away from each other, the tensor $A^{\text{CDL}}$ corresponding to this black square should factorize into the tensor product of four corner matrices $C$. 
A tensor with the structure of $A^{\text{CDL}}$ is called a corner double-line (CDL) tensor.

\begin{figure}[t]
    \includegraphics[scale=0.9,valign=c]{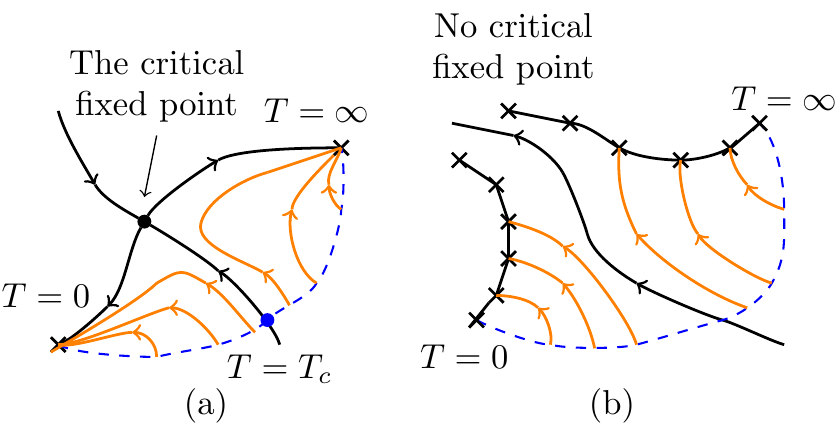}
    \caption{\label{fig:tensorRGflow}
        Schematic RG flows of the 2D Ising model, without and with the problem of local correlations.
        Each point on the dashed line represents the lattice model at a given temperature and is the starting point of an RG transformation.
        The solid lines with arrows represent different RG flows.
        (a) The correct RG flow. There are one $T=0$ fixed point, one $T=\infty$ fixed point and one critical fixed point.
        (b) The RG flows generated by the TRG and the HOTRG\@.
        Due to the problem of local correlations, the two trivial fixed points become two fixed lines, and the critical fixed point disappears.
    }
\end{figure}

The CDL tensors are fixed points of the RG equations of the TRG~\cite{LevinTalk,GuWen2009,tnr,gilts} and the HOTRG~\cite{hotrgfixpoint}.
This shows that the TRG and the HOTRG have difficulty in integrating out the local interactions among the spin variables around the corners. 
If we start with two temperatures $T_1 \neq T_2$, both larger than the critical temperature $T_c$ of the 2D Ising model, either of these two methods will generate tensors flowing to two different CDL tensors $A^{\text{CDL}}_1 \neq A^{\text{CDL}}_2$, as a natural consequence of the fact that these CDL tensors depend, directly, on the bare interaction constants. 
At criticality, the previous numerical calculations indicate that we will never reach a critical fixed-point tensor~\cite{Berker2008,tnr}.
Their calculations suggest tensor RG flows shown in Fig.~\ref{fig:tensorRGflow}(b), where the low- and high-temperature fixed points turn into two fixed lines and the critical fixed point disappears. 
By comparison, the correct RG flow is shown in Fig.~\ref{fig:tensorRGflow}(a).
We will introduce a way to solve the problem of CDL tensors for the HOTRG in Sec.~\ref{sec:gilthotrg}.
%%% Figure: schematic RG flows

The second obstacle that prevents us from achieving Eq.~\eqref{eq:tensorRGAstar} is that the tensor network representation of the partition function in Fig.~\ref{fig:spin2tensor} and Eq.~\eqref{eq:ZbeforeRG} has gauge redundancy.
If two tensors $\tilde{A}$ and $A$ are related through some invertible matrices $S_x,S_y$ by the gauge transformation
\begin{subequations}\label{def:gaugeTrans}
    \begin{align}\label{def:gaugeTransMath}
        \tilde{A}_{ijkl} = \sum_{\substack{m,n\\p,q}} A_{mnpq} \left(S_x
        \right)_{im} \left(S_y \right)_{jn} \left(S_x^{-1}\right)_{pk}
        \left(S_y^{-1}\right)_{ql}, 
    \end{align}
or pictorially as
    \begin{align}\label{def:gaugeTransPic}
        \includegraphics[scale=0.8,valign=c]{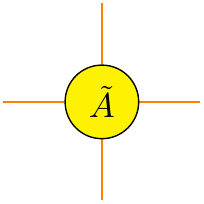}
    =
        \includegraphics[scale=0.8,valign=c]{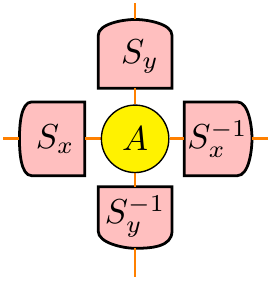},
    \end{align}
\end{subequations}
the two tensor networks formed by $A$ and $\tilde{A}$ represent the same partition function $Z$.
Equation~\eqref{def:gaugeTrans} is a equivalence relation that defines a equivalence class $[A]$.

The gauge redundancy makes the canonical RG prescription in tensor space less straightforward than that in Hamiltonian space.
Even if we have reached a representation tensor $A^*$ of the fixed-point equivalence class $[A^*]$, the coarse graining process could bring this tensor to another representation of $[A^*]$, 
\begin{align}\label{eq:tensorRGAstarnotfix}
    A^* \xrightarrow[\text{graining}]{\text{coarse}} \tilde{A}^*. 
    % \text{ or } 
    %\tilde{A}^* = \mathcal{T}\left(A^* \right).
\end{align}
In general, we must fix the gauge of the tensor during a tensor RG transformation by choosing a preferred set of basis, so that the fixed-point tensor is manifestly fixed, as is shown in Eq.~\eqref{eq:tensorRGAstar}.
We will show how to fix the gauge in Sec.~\ref{sec:gaugefix}.

\subsection{Filtering out local correlations for the HOTRG\label{sec:gilthotrg}}
In this subsection, we present an HOTRG-like scheme to solve the first technical obstacle, the problem of local correlations.
Compared with the state-of-the-art TRG-type methods~\cite{GuWen2009,tnr,tnralgo,tnrplus,looptnr,harada2018,fet,tns,tensor-ring,gilts} that are free of this problem, the proposed scheme can be most easily generalized to 3D and higher and is convenient for the subsequent gauge fixing and linearization procedure.
Specifically, the graph-independent local truncation (GILT)~\cite{gilts} is performed to filter out the problematic local correlations before the coarse graining of the HOTRG\@.
While the GILT may not be the unique solution for removing the local correlations, we adopt it mainly for its conceptual simplicity and ease in adoption.

\begin{figure}[t]
    \includegraphics[scale=0.9,valign=c]{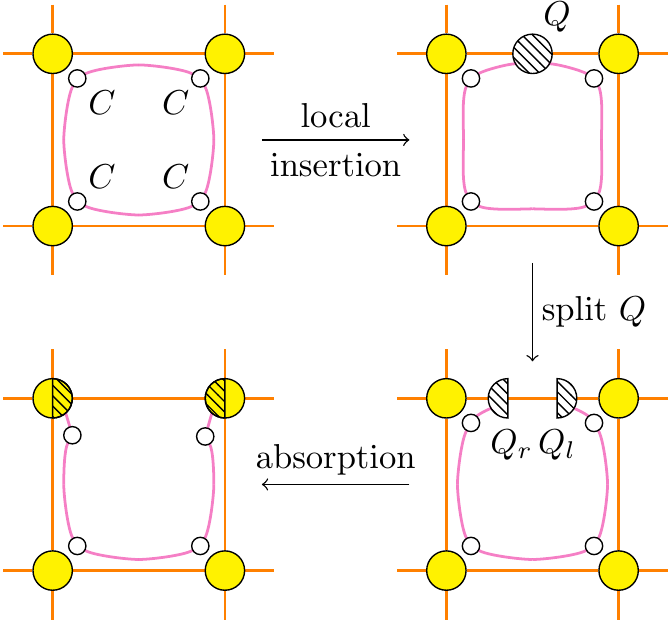}
    \caption{\label{fig:gilt}
        The process of the GILT. 
        Four copies of $C$ matrices are unknown inner structure of the adjacent 4-leg tensors.
        They are drawn explicitly to make the demonstration clearer.
        In the first step, a low-rank matrix $Q$ is inserted into a bond. Then we split $Q$ into two pieces using singular value decomposition. 
        The low-rank matrix $Q$ is constructed so that it cuts the legs of the corner matrices $C$ during the splitting. 
        Finally, the pieces of the matrix $Q$ are absorbed into the two neighboring tensors. 
        The original GILT paper~\cite{gilts} presents a nice way to determine the low-rank matrix $Q$.
        A brief introduction is provided in Appendix~\ref{append:gilthotrg}.
    }
\end{figure}

The key feature of the GILT is that it is a stand-alone procedure to filter out the local correlations and does not change the geometry of a given tensor network, so it is very flexible.
It has been shown that the TRG combined with GILT is able to generate correct tensor RG flows for the 2D Ising model~\cite{gilts} and the 2D $\phi^4$ theory~\cite{Delcamp2020}.
Figure~\ref{fig:gilt} summarizes the basic process of the GILT. 
The loop containing four matrices $C$ inside the plaquette represents the local correlations (see Fig.~\ref{fig:rgschem}(b) and imagine putting four CDL tensors together to form a plaquette). 
The first step, which is the most crucial one, is to insert a low-rank matrix $Q$ into the leg we wish to truncate. 
The tensor network after the insertion should give a good approximation of the initial one.
The remaining two steps are exact. We split $Q$ into two pieces using singular value decomposition and absorb the two pieces into the adjacent two $A$ tensors. 
The bond dimension of the leg is smaller and the local correlations on this leg are filtered out. 

Next, we move on to explain the HOTRG~\cite{hotrg}.
The block-tensor transformation in Eqs.~\eqref{eq:ZunderRG} and~\eqref{eq:exactBlock} is exact but not practical, since the bond dimension grows exponentially in the original lattice size.
The HOTRG is an approximate tensor RG transformation, which can keep the bond dimension from growing.
For the HOTRG in the vertical direction, we aim at the following approximation of a local patch of two copies of $A$ put together vertically,
\begin{align}\label{eq:hotrgProjTrun}
    \includegraphics[scale=0.8,valign=c]{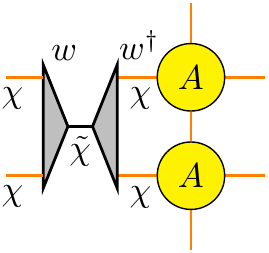}
    \approx 
    \includegraphics[scale=0.8,valign=c]{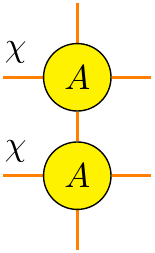},
\end{align}
where $w$ is an isometric tensor to be determined and $w^{\dagger}$ its hermitian conjugate.
The isometry $w$ is a linear mapping: $\mathbb{V}_{\tilde{\chi}} \rightarrow  \mathbb{V}_{\chi}\otimes\mathbb{V}_{\chi}$, where $\mathbb{V}_{\chi}$ denotes a $\chi$-dimensional vector space, and the isometry satisfies $w^{\dagger}w = \mathbb{1}$.
We will later see that the isometric condition of tensor $w$ makes the gauge fixing in the HOTRG easier.
It is shown in Ref.~\cite{hotrg,tnralgo} that a good approximation can be achieved if the isometry $w$ is a collection of $\tilde{\chi}$ eigenvectors corresponding to the first $\tilde{\chi}$ largest eigenvalues of the $\chi^2$-by-$\chi^2$ positive semi-definite matrix $MM^{\dagger}$, with the matrix $M$ defined as
\begin{align}\label{def:M-AA} 
    M = 
    \includegraphics[scale=0.9,valign=c]{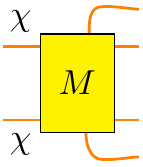}
    \equiv 
    \includegraphics[scale=0.8,valign=c]{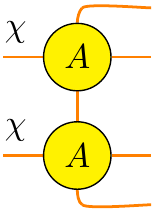}.
\end{align}
We use the approximation in Eq.~\eqref{eq:hotrgProjTrun} to replace all pairs of $A$ tensors in the tensor network representation of the partition function $Z_{4\times4}$ in Eq.~\eqref{eq:ZunderRG} to get
\begin{align}\label{eq:Zapproxy} 
    Z_{4 \times 4}
&\textapproxmy{\eqref{eq:hotrgProjTrun}}
    \includegraphics[scale=1.0,valign=c]{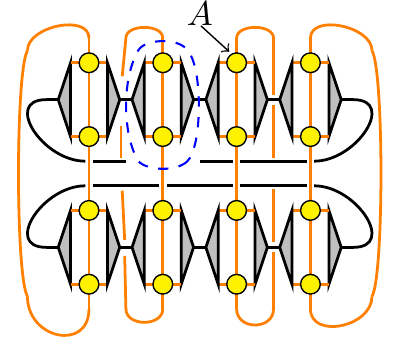}
    \nonumber\\ &= 
    \includegraphics[scale=1.0,valign=c]{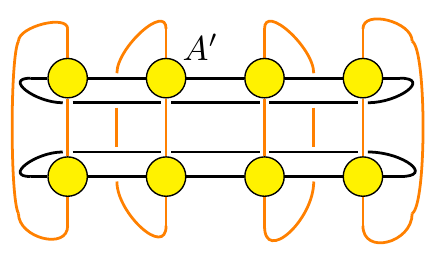},
\end{align}
where in the second step, we contract two $A$ tensors and $w, w^{\dagger}$ in the dashed circle to get a coarser tensor $A'$,
\begin{align}\label{def:Apycontr}
    \includegraphics[scale=1.0,valign=c]{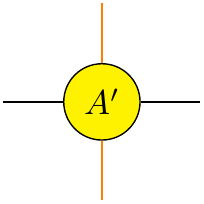}
    \equiv 
    \includegraphics[scale=0.8,valign=c]{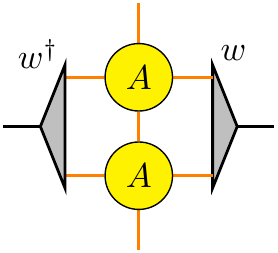}.
\end{align}
Notice in the approximation step in Eq.~\eqref{eq:Zapproxy}, we move the two leftmost $w$ tensors to the right because we have a periodic boundary condition. 
Equation~\eqref{def:Apycontr} defines the HOTRG coarse graining in the vertical direction.
We usually choose $\tilde{\chi} \leq \chi_{\text{max}}$ in Eq.~\eqref{eq:hotrgProjTrun} to prevent the bond dimension from growing.

In Appendix~\ref{append:gilthotrg}, we demonstrate how to choose the plaquettes and where to insert the low-rank matrices in the GILT to filter out the local correlations for the HOTRG\@. 
The resultant coarse graining in the vertical direction looks similar to Eq.~\eqref{def:Apycontr}, only with a few more pieces of the low rank matrices from the GILT inserted into the bonds between the tensor $A$ and the isometric tensors $w, w^{\dagger}$.
The two coarse-graining steps in both vertical and horizontal directions together define the RG equation of this HOTRG-like scheme (before the gauge fixing),
\begin{align}\label{def:RGeqGiltHOTRG}
    \includegraphics[scale=1.0,valign=c]{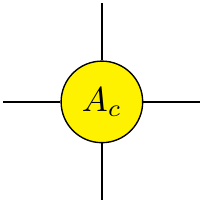}
    =
    \includegraphics[scale=0.8,valign=c]{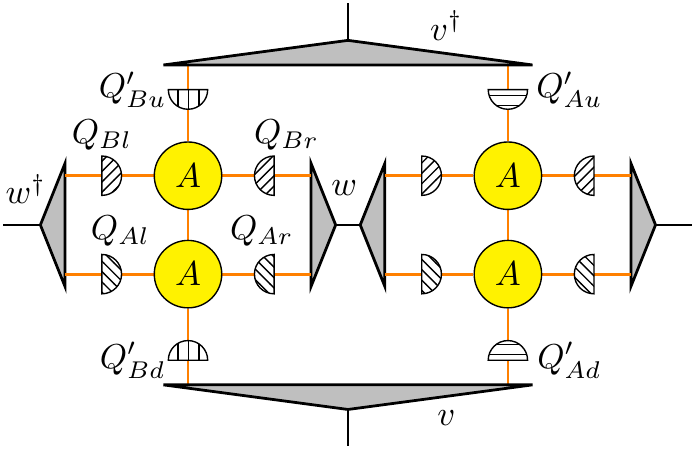}.
\end{align}
The computational costs of the determination of all the isometric tensors and low-rank matrices are $O(\chi^6)$, while those of the contraction of the tensor network on the right hand side of Eq.~\eqref{def:RGeqGiltHOTRG} are $O(\chi^7)$.
The computation costs of this HOTRG-like coarse graining are thus $O(\chi^7)$, the same as the original HOTRG\@.

The coarse graining defined in Eq.~\eqref{def:RGeqGiltHOTRG} is able to simplify the $A^{\text{CDL}}$ tensor in Fig.~\ref{fig:rgschem}(b) to a single number, 
\begin{align}\label{eq:CDL2number}
    \includegraphics[scale=1.0,valign=c]{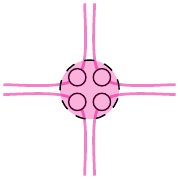}
    \xrightarrow{\text{Eq.~\eqref{def:RGeqGiltHOTRG}}}
    \left(
        \includegraphics[scale=0.8,valign=c]{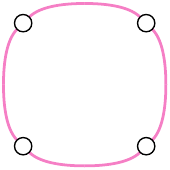}
    \right)^4.
\end{align}
Equation~\eqref{eq:CDL2number} shows that this HOTRG-like  scheme can successfully filter out the local correlations among the spin variables around the corners at the lattice scale (see Fig.~\ref{fig:rgschem}(b)).
Since the CDL tensors are no longer fixed points for the RG equation of this HOTRG-like scheme, the peculiar fixed lines in Fig.~\ref{fig:tensorRGflow}(b) generated by the HOTRG will collapse to fixed points; we expect that Eq.~\eqref{def:RGeqGiltHOTRG} is able to exhibit the critical fixed point tensor shown schematically in Fig.~\ref{fig:tensorRGflow}(a).
%
%%% Figure: summary of the GILT

\subsection{Gauge fixing and the linearized tensor renormalization group transformation\label{sec:gaugefix}}
We show how the gauge is fixed and give the explicit expression of the linearized RG equation of the HOTRG-like scheme in this subsection.

Part of the gauge can be fixed if the physical model possesses a global internal symmetry.
The global symmetry can be incorporated into the tensor network representation of the model~\cite{Singh2010SymTen,Singh2011U1Ten,Singh2012SU2Ten}; it is a generalization of Schur's lemma from matrices to general tensors.
For the 2D Ising model, $\mathbb{Z}_2$ symmetry can be imposed.
Each index of the tensor $A$ breaks into even and odd sectors.
Half of the gauge is fixed since $A$ is in the basis where the states in the even sector transform trivially and
the states in the odd sector is multiplied by $-1$ under the spin flip operation.

Most of the remaining gauge in the degenerate sectors of $A$ can be fixed by going to the diagonal basis of the tensor.
We show how the $S_x$ gauge redundancy in Eq.~\eqref{def:gaugeTrans} is fixed. The $S_y$ one can be dealt with in the same way.
Given a tensor $A$, we first contract its two vertical legs to produce a transfer matrix $N_x$,
\begin{align}\label{def:Nx}
    \includegraphics[scale=0.8,valign=c]{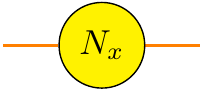}
    = 
    \includegraphics[scale=0.8,valign=c]{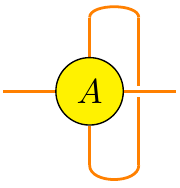}.
\end{align}
We then find the eigenvalue decomposition of this matrix,
\begin{align}\label{eq:eigdcpNx}
    \includegraphics[scale=0.8,valign=c]{Nx.pdf}
    = 
    \includegraphics[scale=0.8,valign=c]{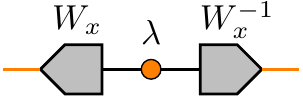}\text{ },
\end{align}
where $\lambda$ is the diagonal matrix encoding eigenvalues.
The gauge fixing transformation in the horizontal direction is defined by acting the invertible matrix $W_x$ and its inverse on the horizontal legs of the tensor $A$,
\begin{align}\label{def:gaugefixHori}
    \includegraphics[scale=0.8,valign=c]{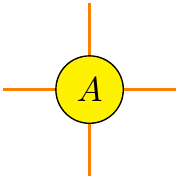}
    \xrightarrow[\text{gauge fixing}]{\text{horizontal}} 
    \includegraphics[scale=0.8,valign=c]{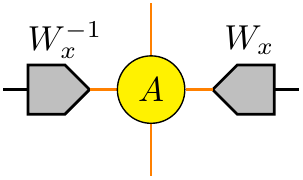}.
\end{align}
It is shown in Appendix~\ref{append:proofgaugefix} that the above procedure fully fixes the gauge redundancy in two horizontal legs except the phase ambiguities if there is no degeneracy in the spectrum of $N_x$. 

The gauge fixing procedure described in Eqs.~\eqref{def:Nx} to~\eqref{def:gaugefixHori} is general for all TRG-type techniques.
However, this procedure is not necessary for the HOTRG-like scheme applied to systems with spatial reflection symmetries like the 2D Ising model, since the RG equation in Eq.~\eqref{def:RGeqGiltHOTRG} has a preferred set of basis.
As a result, the gauge redundancy in Eq.~\eqref{def:gaugeTrans} collapses into phase ambiguities (or sign ambiguities for real tensors) in the HOTRG-like scheme\@.
To make things as simple as possible, we focus on real tensors in the following discussions.
The generalization to complex tensors is straightforward.
%

% Write the tensor RG equation in Eq.~\eqref{def:RGeqGiltHOTRG} schematically as $A_c = \mathcal{T}\left(A\right)$.
For two real tensors $A, \tilde{A}$ that are related by the gauge transformation defined in Eq.~\eqref{def:gaugeTrans} where we further restrict $S_x,S_y$ to be orthogonal matrices, the new tensors generated by Eq.~\eqref{def:RGeqGiltHOTRG}, $A_c$ and $ \tilde{A}_c$, are equal up to sign ambiguities,
\begin{align}\label{eq:signAmbi}
    \left(\tilde{A}_c \right)_{ijkl} =
    \left(A_c\right)_{ijkl}(d_x)_i (d_y)_j (d_x)_k (d_y)_l, 
\end{align}
where $d_x,d_y$ are vectors with components $\pm 1$. 
The proof of the property of the HOTRG-like scheme in Eq.~\eqref{eq:signAmbi} is provided in Appendix~\ref{append:proofgaugefix}.
Imagine that we manage to fix the sign ambiguities, then we can write Eq.~\eqref{def:RGeqGiltHOTRG} after the gauge fixing schematically as $A_c = \mathcal{T}(A)$. The RG equation of the HOTRG-like scheme ensures
\begin{align}\label{eq:HOTRGgaugefix}
    \mathcal{T}(A) = \mathcal{T}(\tilde{A}).
\end{align}
Since the orthogonal matrices $S_x, S_y$ are arbitrary, equation~\eqref{eq:HOTRGgaugefix} says that the whole equivalence class $[A]$ will be mapped into the same tensor $A_c$. 
This means that the HOTRG-like scheme, after incorporating the sign fixing step, will choose a preferred set of basis. 
It is worth to mention that the TRG has a similar property~\cite{kadanoff2014}.
For a fixed-point tensor, equation~\eqref{eq:HOTRGgaugefix} indicates that we can start with any representation $\tilde{A}^*$ of the equivalence class $[A^*]$, and the HOTRG-like scheme will bring $\tilde{A}^*$ to the proper basis; further coarse graining will satisfy Eq.~\eqref{eq:tensorRGAstar},
\begin{align}\label{eq:GiltHOTRGfixT}
    \mathcal{T}(\tilde{A}^*) =
    \mathcal{T}\left(\mathcal{T}(\tilde{A}^*)  \right) \equiv A^*.
\end{align}

The sign ambiguities $d_x,d_y$ in Eq.~\eqref{eq:signAmbi} can be determined by comparing the sign of the components of $\tilde{A}_c$ and $A_c$.
For example, upon making sure $(\tilde{A}_c)_{1111}$ and $(A_c)_{1111}$ are both positive, set $j = k = l = 1$ in Eq.~\eqref{eq:signAmbi} to have
\begin{align}\label{eq:finddx}
    (\tilde{A}_c)_{i111} =
    (A_c)_{i111}(d_x)_i. 
\end{align}
The relative sign of $(\tilde{A}_c)_{i111}$ and $(A_c)_{i111}$ determines $(d_x)_i$.
However, this sign fixing method breaks down if both $(\tilde{A}_c)_{i111}$ and $(A_c)_{i111}$ vanish, which occurs as long as there is a symmetry.
This is the reason why we first fix part of the gauge by exploiting the global internal symmetry of the physical model.
Then, we can apply Eq.~\eqref{eq:finddx} in each degenerate sector of the tensor.
The detailed implementation of the sign fixing procedure for $\mathbb{Z}_2$ symmetric tensors can be found in the source code of this paper (see Appendix~\ref{append:sc}).

After reaching the fixed-point tensor $A^*$ in Eq.~\eqref{eq:GiltHOTRGfixT}, the next step is to linearize the tensor RG equation in Eq.~\eqref{def:RGeqGiltHOTRG}.
We substitute $A = A^* + \delta A$ into the right hand side of Eq.~\eqref{def:RGeqGiltHOTRG} and collect terms that are first order in $\delta A$ to get $\delta A_c$,
\begin{widetext}
    \begin{align}\label{eq:respMatGiltHOTRG}
    \includegraphics[scale=1.0,valign=c]{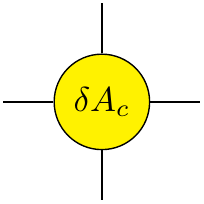}
    &=
    \includegraphics[scale=0.8,valign=c]{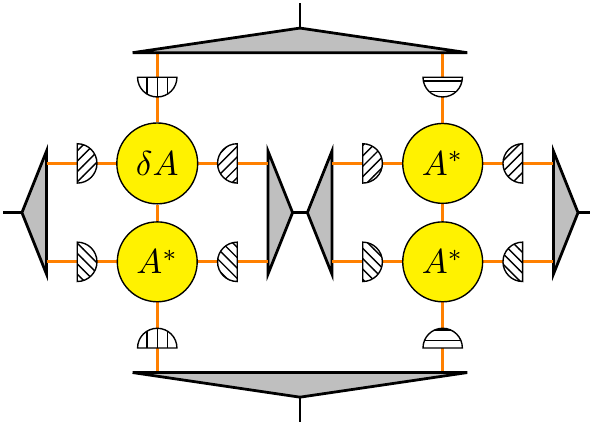}
    +
    \includegraphics[scale=0.8,valign=c]{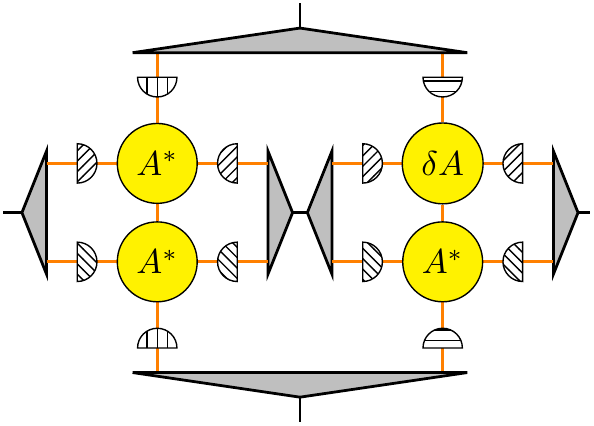}
    + \text{ two similar terms},
    \end{align}
\end{widetext}
where we refrain from drawing $d_x,d_y$ coming from the sign fixing procedure.
The result resembles the product rule for taking the differentials in calculus.
Equation~\eqref{eq:respMatGiltHOTRG} provides a simple pictorial representation of the linearized tensor RG equation $\mathcal{R}$ in Fig.~\ref{fig:linearedRGschem} and Eq.~\eqref{eq:respMatTen} for the HOTRG-like scheme.
The computational costs of the contraction of the right hand side of Eq.~\eqref{eq:respMatGiltHOTRG} are $O(\chi^7)$, the same as those of the HOTRG-like coarse graining in Eq.~\eqref{def:RGeqGiltHOTRG}.
In practice, after the fixed-point tensor $A^*$, the pieces of low-rank matrices $Q_A,Q_B$ and the isometric tensors $w, v$ in Eq.~\eqref{def:RGeqGiltHOTRG} are determined, automatic differentiation can linearize Eq.~\eqref{def:RGeqGiltHOTRG} around $A^*$ and generate Eq.~\eqref{eq:respMatGiltHOTRG} for us.
There are many libraries that support automatic differentiation, including PyTorch~\cite{pytorch} and JAX~\cite{jax2018github}.

\section{Examples\label{benchmark}}
We use the classical Ising model in 1D and 2D to demonstrate how to carry out the canonical RG prescription in tensor space. 
The Ising model in 1D serves as a concrete example to elucidate the general argument in Sec.~\ref{RGprescrip}. 
The Ising model in 2D provides more nontrivial benchmark results for our method.

\subsection{The Ising Model in 1D\label{benchmark:1DIsing}}
The Ising model in 1D has an exact real-space RG transformation realized via decimation. 
Even better, the decimation has a natural tensor network representation. 
This makes the Ising model in 1D a nice example to see the relation between the old and the new approaches of the canonical RG prescription.

\begin{figure}[t]
    \includegraphics[width=0.9\columnwidth,valign=c]{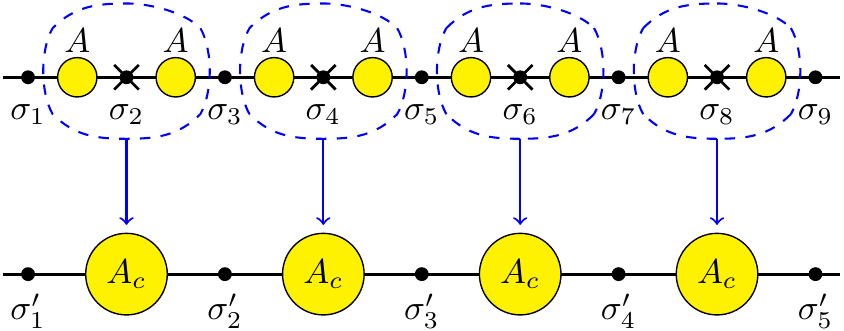}
    \caption{\label{fig:Ising1D-decimation}
        The decimation for the 1D Ising model. 
        The black dots are spin variables. 
        The spins on even sites $\sigma_2,\sigma_4,\ldots$ are summed over and the remaining spins $\sigma_1,\sigma_3,\ldots$ are renamed $\sigma_1',\sigma_2',\ldots$ to become new spin variables. 
        In the tensor network language, this decimation is nothing but a matrix multiplication of two transfer matrices to form a coarse-grained matrix $A_c = AA$.
    }
\end{figure}

The partition function is
\begin{align}\label{def:Z4Ising1D}
    Z_{\text{1D}} = \sum_{\{\sigma_j \} } \exp{\left[\sum_{i=1}^N
    \mathscr{H}\left(\sigma_i,\sigma_{i+1}\right)  \right]},
\end{align}
where the local interactions involve the nearest-neighbor term at most
\begin{align}\label{def:H4Ising1D}
    \mathscr{H}\left(\sigma_1, \sigma_2\right) = g +
    \frac{h}{2}\left(\sigma_1 + \sigma_2\right) + K\sigma_1 \sigma_2.
\end{align}
The decimation process is shown in Fig.~\ref{fig:Ising1D-decimation}.
It is realized by summing over all the even-numbered spins and then renumber the remaining odd-numbered spins. 
%
%% Figure: decimation of the 1D Ising model
We denote $\sigma_i'=\sigma_{2i-1}, s_i = \sigma_{2i}$ and sum over all $s$-spins in the partition function in Eq.~\eqref{def:Z4Ising1D} to have
\begin{align}\label{eq:oldK2newKZ}
    Z_{\text{1D}} = 
    \sum_{\{\sigma_j'\}} \sum_{\{s_j \}}
    \exp{\left[ \sum_i^{N/2} \left[\mathscr{H}\left(\sigma_i',s_i\right)
    + \mathscr{H}\left(s_i,\sigma_{i+1}'\right)\right]\right]},
\end{align}
from which we can define the effective local interaction $\mathscr{H}'$ through
\begin{align}\label{eq:oldK2newK}
    \exp{\left[\mathscr{H}'\left(\sigma_1',\sigma_2'\right)\right]} =
    \sum_{s=\pm 1}\exp{\left[\mathscr{H}\left(\sigma_1',s\right) +
        \mathscr{H}\left(s,\sigma_2'\right)\right]},
\end{align}
where the effective local interaction has the same form as the old one in Eq.~\eqref{def:H4Ising1D} but with new coupling constants $g',h',K'$,
\begin{align}\label{def:newH4Ising1D}
    \mathscr{H}'\left(\sigma_1,\sigma_2\right) = g' +
    \frac{h'}{2}\left(\sigma_1 + \sigma_2\right) + K' \sigma_1 \sigma_2.
\end{align}
The partition function can be fully described by the new $\sigma'$-spins,
\begin{align}\label{eq:Z2Ising1Dnew}
    Z_{\text{1D}} = \sum_{\{\sigma_j'\}}
    \exp{\left[\sum_{i=1}^{N/2}\mathscr{H}'\left(\sigma_i',\sigma_{i+1}'\right)\right]}.
\end{align}
Equations~\eqref{def:H4Ising1D}, \eqref{eq:oldK2newK} and~\eqref{def:newH4Ising1D} together define the RG equation that maps the old coupling constants $(g,h,K)$ to the new coupling constants $(g',h',K')$. 
The explicit expression of the RG equation can be found in Kardar's textbook~\cite{kardar2007}. 
The RG equation has two fixed points, one for high-temperature phase and the other for low-temperature phase. 
Let us focus on the high-temperature fixed point here, where the coupling constants are $g^* = \log\left(1/2\right),h^*=0,K^*=0$. 
The linearized RG equation around this fixed point gives $\delta g' = 2\delta g, \delta h' = \delta h, \delta K' = 0\times \delta K$. 
The matrix $\mathcal{R}$ is in its diagonal form with eigenvalues $2,1,0$ for $\delta g,\delta h,\delta K$ respectively.

Next, we translate the above decimation process into tensor network language. 
We first define the tensor $A$ sitting on the bond connecting two spins shown in Fig.~\ref{fig:Ising1D-decimation} as
\begin{subequations}\label{def:A4Ising1D}
    \begin{align}\label{def:A4Ising1DCompo}
    A_{\sigma_1 \sigma_2} =
    \exp{\left[\mathscr{H}\left(\sigma_1,\sigma_2\right)\right]}.
    \end{align}
    After using the expression for $\mathscr{H}$ in
    Eq.~\eqref{def:H4Ising1D}, we have
    \begin{align}\label{def:A4Ising1Depl}
        A = 
    \begin{pmatrix}
    \exp{\left(g + h + K\right)} & \exp{\left(g - K\right)} \\
    \exp{\left(g - K\right)} & \exp{\left(g - h + K\right)} \\
    \end{pmatrix},
    \end{align}
\end{subequations}
which is the familiar transfer matrix. Each component of the tensor $A$ is a function of coupling constants $g, h, K$, as is claimed in Eq.~\eqref{eq:K2A}. 
The partition function in Eq.~\eqref{def:Z4Ising1D} can be rewritten as
\begin{align}\label{eq:Z4Ising1DbyA}
    Z_{\text{1D}} = \sum_{\{\sigma_j\}} \bigotimes_{i=1}^N A_{\sigma_i
        \sigma_{i+1}}.
\end{align}
The decimation in the tensor network language is a multiplication of two old $A$ matrices to form a new $A_c$ matrix,
\begin{align}\label{def:Ising1DRGeqTen}
    A_c = AA.
\end{align}
In terms of the new $A_c$ matrix, the partition function is
\begin{align}\label{eq:Z4Ising1DbyAp}
    Z_{\text{1D}} = \sum_{\{\sigma_j'\}} \bigotimes_{i=1}^{N/2}
    (A_c)_{\sigma_i' \sigma_{i+1}'}.
\end{align}
Equation~\eqref{def:Ising1DRGeqTen} is the RG equation in the tensor network language. 
Now, each component of $A_c$ is a function of coupling constants $g,h,K$ but with different functional form, as is claimed in Eq.~\eqref{eq:tensorEleRG}. 
If we further require that $A_c$ should have the same form as $A$ in Eq.~\eqref{def:A4Ising1D} but with $\mathscr{H}$ replaced by $\mathscr{H}'$, new coupling constants $g',h',K'$ can be solved in terms of the old ones, which is what we do in the conventional approach. 
The advantage of using the tensor network language is that the RG equation in Eq.~\eqref{def:Ising1DRGeqTen} suffices for the canonical RG prescription in tensor space. 
First, let us set the coupling constants in Eq.~\eqref{def:A4Ising1Depl} to be the high-temperature fixed point $g^* = \log{\left(1/2\right)}, h^*=0, K^* = 0$ to get the fixed-point tensor,
\begin{align}\label{eq:fixedA4Ising1D}
    A^* = \frac{1}{2}
\begin{pmatrix}
    1 & 1 \\
    1 & 1 \\
\end{pmatrix}.
\end{align}
It can be checked that $A^* A^* = A^*$. 
The linearized version of Eq.~\eqref{def:Ising1DRGeqTen} around this fixed-point tensor is
\begin{align}\label{eq:Ising1DRespEq}
    \delta A_c = \delta A A^* + A^* \delta A = I \delta A A^* + A^*
    \delta A I,
\end{align}
where in the last equal sign, we add two identity matrices. 
Write Eq.~\eqref{eq:Ising1DRespEq} in its component form, we have $\left(\delta A_c\right)_{ab} = \sum_{\alpha,\beta}I_{a\alpha}\left(\delta A\right)_{\alpha\beta} \left(A^*\right)_{\beta b} + \left(A^*\right)_{a\alpha} \left(\delta A\right)_{\alpha \beta} I_{\beta b}$. 
We can read off the matrix of the linearized RG equation as
\begin{align}\label{eq:Ising1DRespMat}
    \mathcal{R}_{(ab)(\alpha \beta)} = \frac{\left(\delta
    A_c\right)_{ab}}{\left(\delta A\right)_{\alpha \beta}} =
    I_{a\alpha}\left(A^*\right)_{\beta b} + \left(A^*\right)_{a\alpha}
    I_{\beta b},
\end{align}
where we group two indices $a,b$ as a single index $(ab)$, and $\alpha,\beta$ as $(\alpha\beta)$. 
If we put the grouped index into the following order,
\begin{align}\label{def:orderConvention}
    (11) \rightarrow 1, (12) \rightarrow 2, (21) \rightarrow 3, (22)
    \rightarrow 4,
\end{align}
the matrix takes the following value
\begin{align}\label{eq:Ising1DRespMatNum}
    \mathcal{R} = 
\begin{pmatrix}
    1 & 1/2 & 1/2 & 0 \\
    1/2 & 1 & 0 & 1/2 \\
    1/2 & 0 & 1 & 1/2 \\
    0 & 1/2 & 1/2 & 1 \\
\end{pmatrix}.
\end{align}
This matrix $\mathcal{R}$ in Eq.~\eqref{eq:Ising1DRespMatNum} is a symmetric, and we can find its eigenvalues and eigenvectors: $\lambda_1 = 2,\mathbf{v}_1 = (1,1,1,1)^T$; $\lambda_2 = 1,\mathbf{v}_2 = (1,0,0,-1)^T$; $\lambda_3 =1, \mathbf{v}_3 = (0,1,-1,0)^T$ and $\lambda_4 = 0, \mathbf{v}_4 = (1,-1,-1,1)^T$. 
The eigenvalues are the same as what we get in the conventional method. 

The relation between the canonical RG prescription in tensor space and the Hamiltonian space can be clarified by noticing that the relation between the coupling constants and the tensor $A$ is given in Eq.~\eqref{def:A4Ising1Depl}. 
We perturb the coupling constants around the fixed point, $g_p = \log{(1/2)} + \delta g, h_p = \delta h, K_p = \delta K$, substitute them into the right hand side of Eq.~\eqref{def:A4Ising1Depl} and Taylor expand to get the perturbed tensor,
\begin{align}\label{eq:Apert4Ising1D}
    A_p &= A^* + \frac{1}{2} \delta g
    \begin{pmatrix}
    1 & 1 \\
    1 & 1 \\
    \end{pmatrix}
    + \frac{1}{2} \delta h
    \begin{pmatrix}
    1 & 0 \\
    0 & -1 \\
    \end{pmatrix} \nonumber\\
    &+ \frac{1}{2} \delta K
    \begin{pmatrix}
    1 & -1 \\
    -1 & 1 \\
    \end{pmatrix}
    + \text{ higher-order terms }.
\end{align}
We can read off $\delta A = A_p - A^*$ as
\begin{align}\label{eq:deltaA4Ising1D}
    \delta A = \frac{1}{2} \delta g
    \begin{pmatrix}
    1 & 1 \\
    1 & 1 \\
    \end{pmatrix}
    + \frac{1}{2} \delta h
    \begin{pmatrix}
    1 & 0 \\
    0 & -1 \\
    \end{pmatrix} 
    + \frac{1}{2} \delta K
    \begin{pmatrix}
    1 & -1 \\
    -1 & 1 \\
    \end{pmatrix},
\end{align}
which is Eq.~\eqref{eq:deltaK2deltaA} in practice. 
Recall the order convention in Eq.~\eqref{def:orderConvention}, we see the correspondence $\mathbf{v}_1 \leftrightarrow \delta g$, $\mathbf{v}_2 \leftrightarrow \delta h$ and $\mathbf{v}_4 \leftrightarrow \delta K$.

\subsection{The Ising Model in 2D\label{benchmark:2DIsing}}
There is no exact RG transformation for the Ising model in 2D, so we will use the HOTRG-like scheme developed in Sec.~\ref{sec:gilthotrg} to generate RG flows in tensor space. 
The source code of the calculations in this subsection, including the implementations of the HOTRG-like scheme and its linearized version, can be found in Appendix~\ref{append:sc}.

The partition function is given in Eq.~\eqref{eq:2DIsingZ} and we translate the partition function into a tensor network in Fig.~\ref{fig:spin2tensor}. 
Let us denote the initial tensor in Eq.~\eqref{def:tensorA} as $A^{(0)}$. 
To prevent a rapid grow of the magnitude of the tensor during the RG transformation, we pull out the Frobenius norm of the tensor, $A^{(0)} = \Vert A^{(0)}\Vert \mathcal{A}^{(0)}$, to define a normalized tensor $\mathcal{A}^{(0)}$. 
The normalized tensor $\mathcal{A}^{(0)}$ will be fed into the RG equation of the HOTRG-like scheme in Eq.~\eqref{def:RGeqGiltHOTRG} and we denote the output coarse-grained tensor as $A^{(1)}$, from which the norm $\Vert A^{(1)}\Vert$ is pulled out and the normalized tensor $\mathcal{A}^{(1)}$ is defined the same way as the previous step. 
The process can be repeated so we will have $A^{(n)} = \Vert A^{(n)}\Vert \mathcal{A}^{(n)}$ at the $n$-th step. 
The RG flow in tensor space can be conveniently visualized by examining the evolution of the norms $\Vert A^{(n)}\Vert$ as the RG step $n$ increases.

\begin{figure}[t]
    \includegraphics[width=\columnwidth,valign=c]{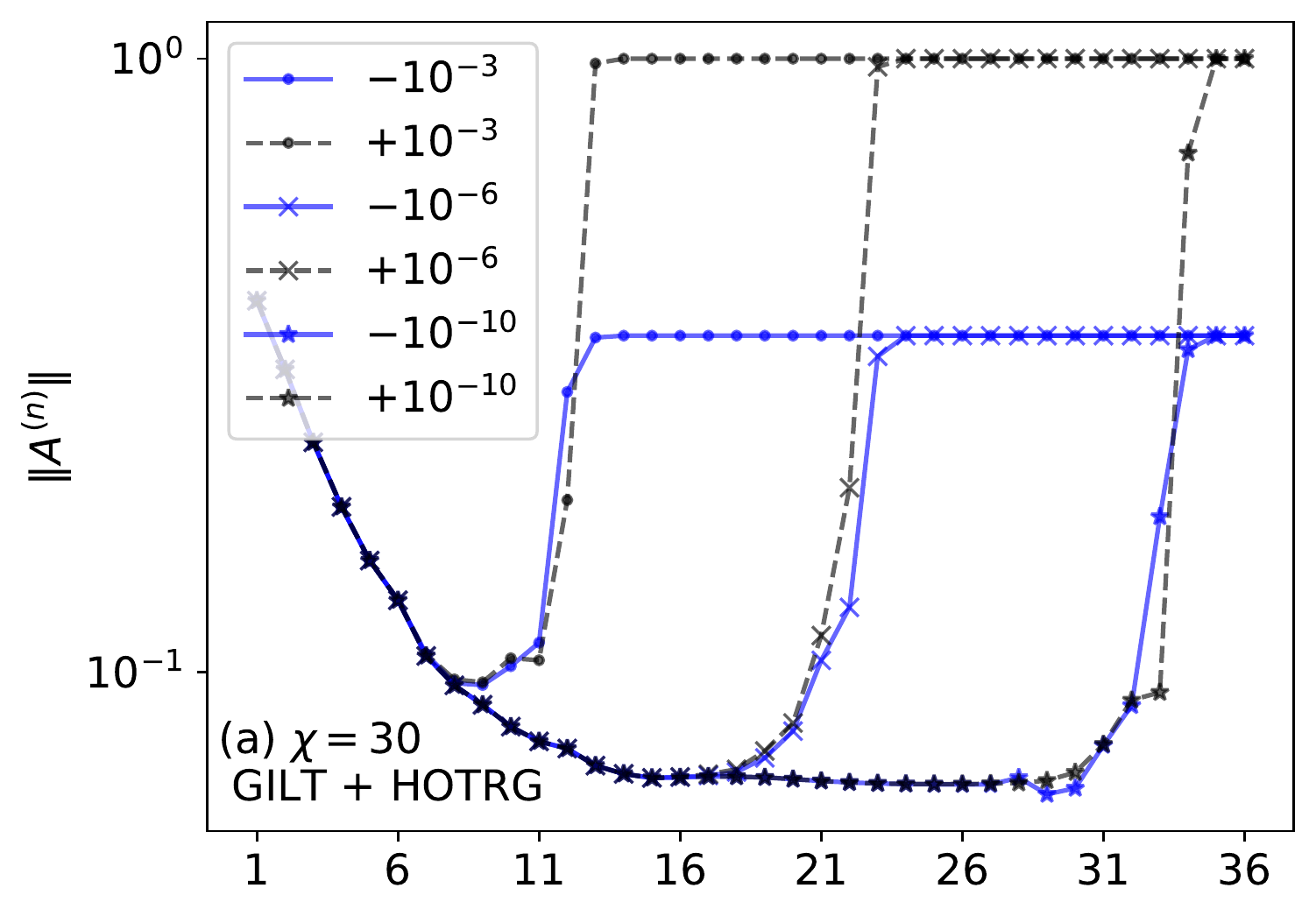}
    \includegraphics[width=\columnwidth,valign=c]{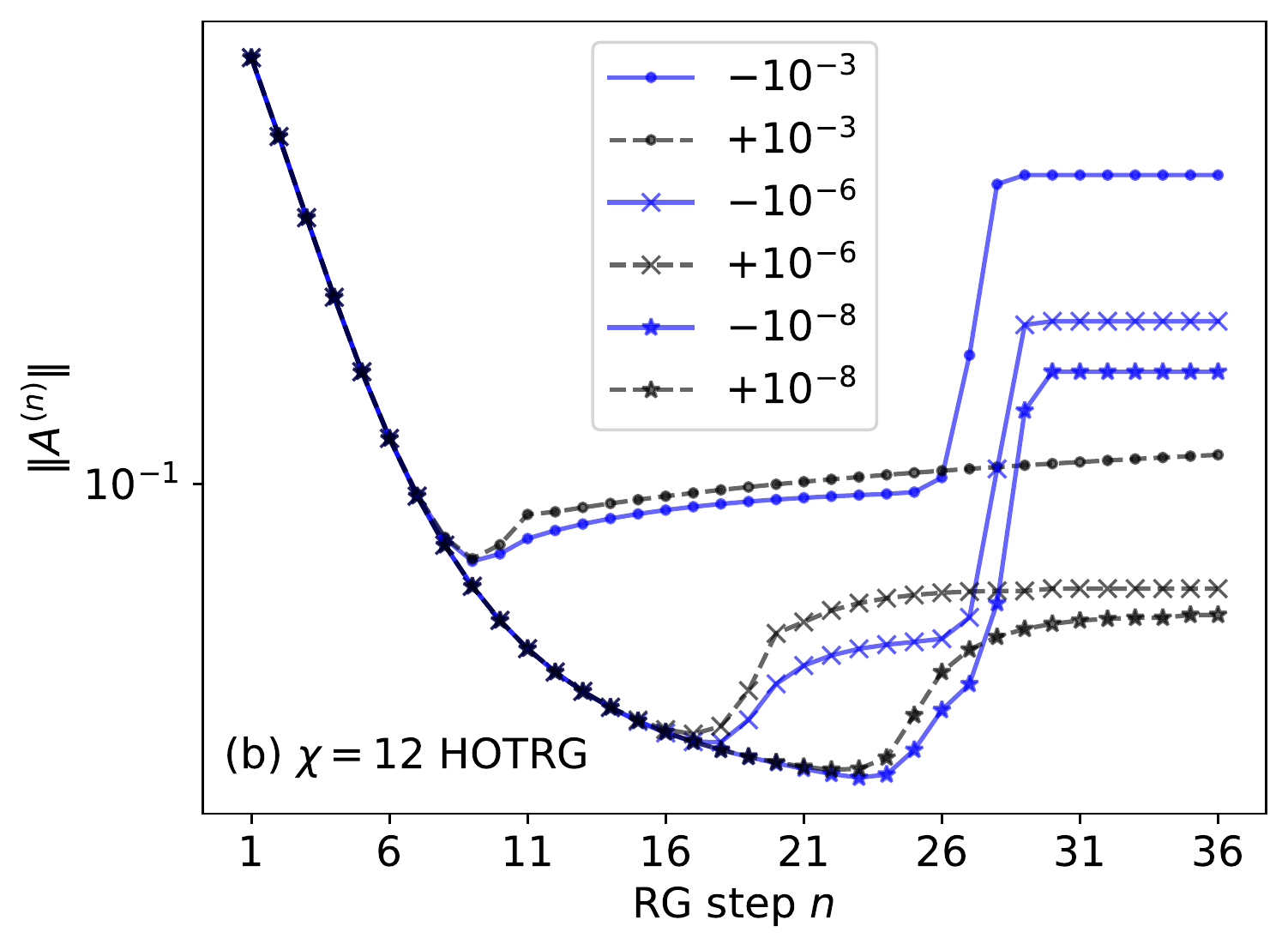}
    \caption{\label{fig:flowAnorm}
        The RG flows of the tensor norms $\Vert A^{(n)} \Vert$ at temperatures near the estimated critical temperature $T_c^{[\chi]}$. 
        Different markers represent different deviations $| \Delta T| $ from $T_c^{[\chi]}$. Blue solid lines are for $\Delta T<0$ and black dashed lines for $\Delta T>0$. 
        (a) For the proposed HOTRG-like scheme with $\chi = 30,\epsilon_{\text{gilt}} = 6\times 10^{-6}$, two trivial fixed points are isolated and the critical fixed point can be reached. It corresponds to the schematic RG flows in Fig.~\ref{fig:tensorRGflow}(a). 
        (b) For the plain HOTRG with $\chi = 12$, we have fixed lines and there is no exhibition of a critical fixed point. 
        It corresponds to the schematic RG flows in Fig.~\ref{fig:tensorRGflow}(b).
    }
\end{figure}

The RG flows of the norms $\Vert A^{(n)} \Vert$ indicate the proposed HOTRG-like scheme is capable of generating a correct RG flow for the 2D Ising model in tensor space shown schematically in Fig.~\ref{fig:tensorRGflow}(a). 
For example, for bond dimension $\chi = 30$ and the hyper-parameter of the GILT process $\epsilon_{\text{gilt}} = 6\times 10^{-6}$, Fig.~\ref{fig:flowAnorm}(a) shows several RG flows of the tensor norms $\Vert A^{(n)} \Vert$ at different temperatures. 
For a given bond dimension $\chi$, there is an estimated critical temperature $T_c^{[\chi]}$ at which the tensor hits the critical surface and will flow to the critical fixed-point tensor $(A^{[\chi]})^*_{\text{cr}}$. 
The $T_c^{[\chi]}$ can be determined using the bisection method; for $\chi = 30$, the difference between the estimated value $T_c^{[30]}$and the exact $T_c$, $|T_c^{[30]} - T_c|$, is of order $10^{-6}$. 
At temperatures off by $\Delta T = \pm 10^{-3}$ from $T_c^{[30]}$, the tensor flows to the high- and low-temperature trivial fixed-point tensors respectively before it comes near to $(A^{[30]})^*_{\text{cr}}$. 
As $|\Delta T|$ becomes smaller to order of $10^{-6}$, the tensor will stay in the vicinity of the critical fixed-point tensor $(A^{[30]})^*_{\text{cr}}$ for a while and then flow away to one of the two trivial fixed-point tensors. 
If $|\Delta T|$ becomes smaller further to $10^{-10}$, the tensor will stay longer near $(A^{[30]})^*_{\text{cr}}$. 
By comparison, the RG flow of $\Vert A^{(n)}\Vert$ generated by the HOTRG with bond dimension $\chi = 12$ is displayed in Fig.~\ref{fig:flowAnorm}(b)\footnote{
    In principle, we can choose $\chi = 30$ here. 
    In practice, however, our calculations show that the problem of local correlations in the HOTRG becomes worse at larger bond dimensions (see Ref.~\cite{Berker2008} for a similar observation in the context of the TRG), and that $\chi = 12$ is enough to demonstrate this problem.
}. 
The RG flow shows that the HOTRG has difficulty in exhibiting a critical fixed-point tensor or producing isolated trivial fixed-point tensors. 
It is interesting to mention that the RG flow generated by the TRG has a similar behavior~\cite{Berker2008} for bond dimensions $\chi > 8$.
%
%%% Figure: the RG flows of the tensor norms

\begin{figure}[tb]
    \includegraphics[width=\columnwidth]{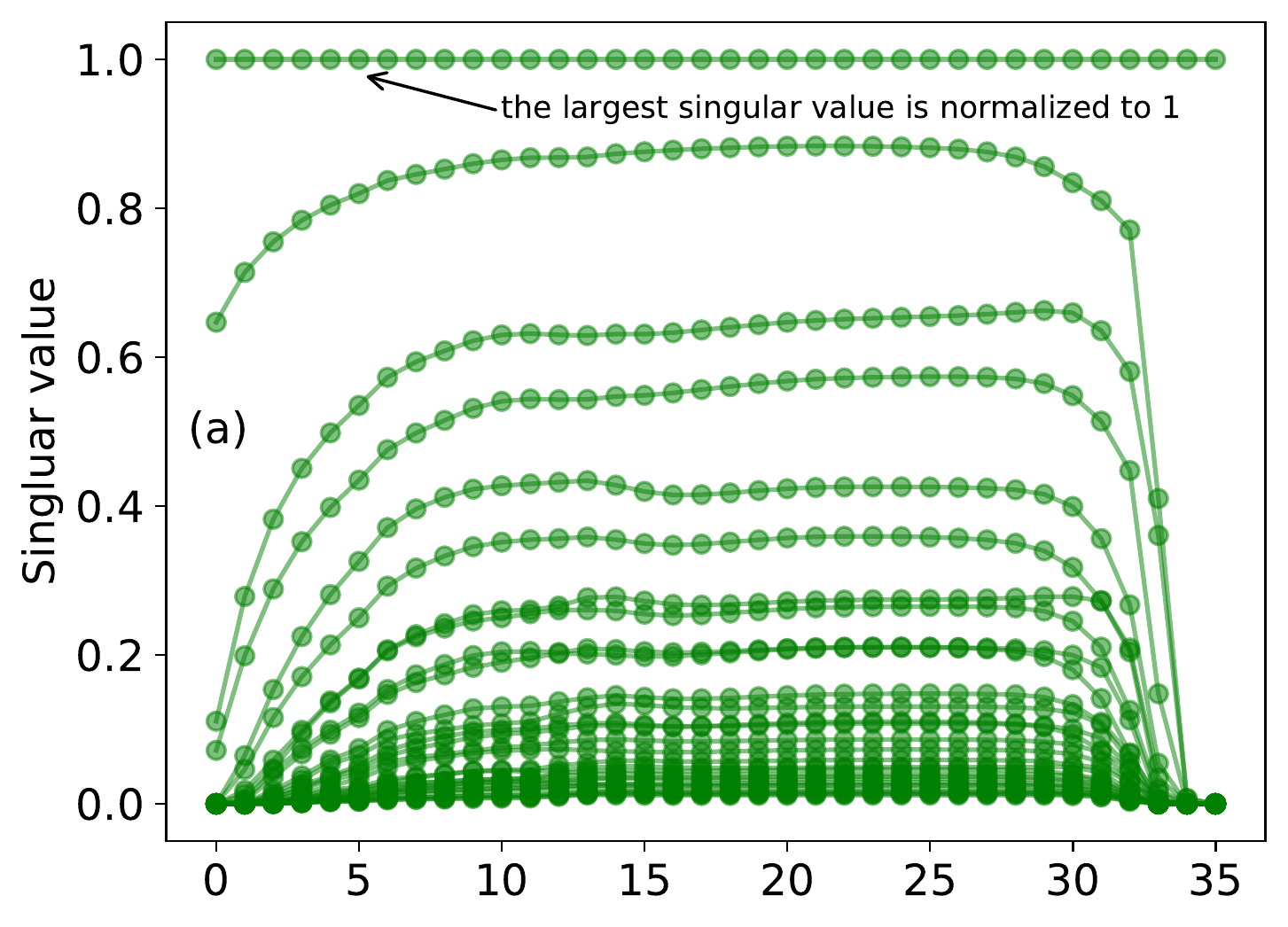}
    \includegraphics[width=\columnwidth]{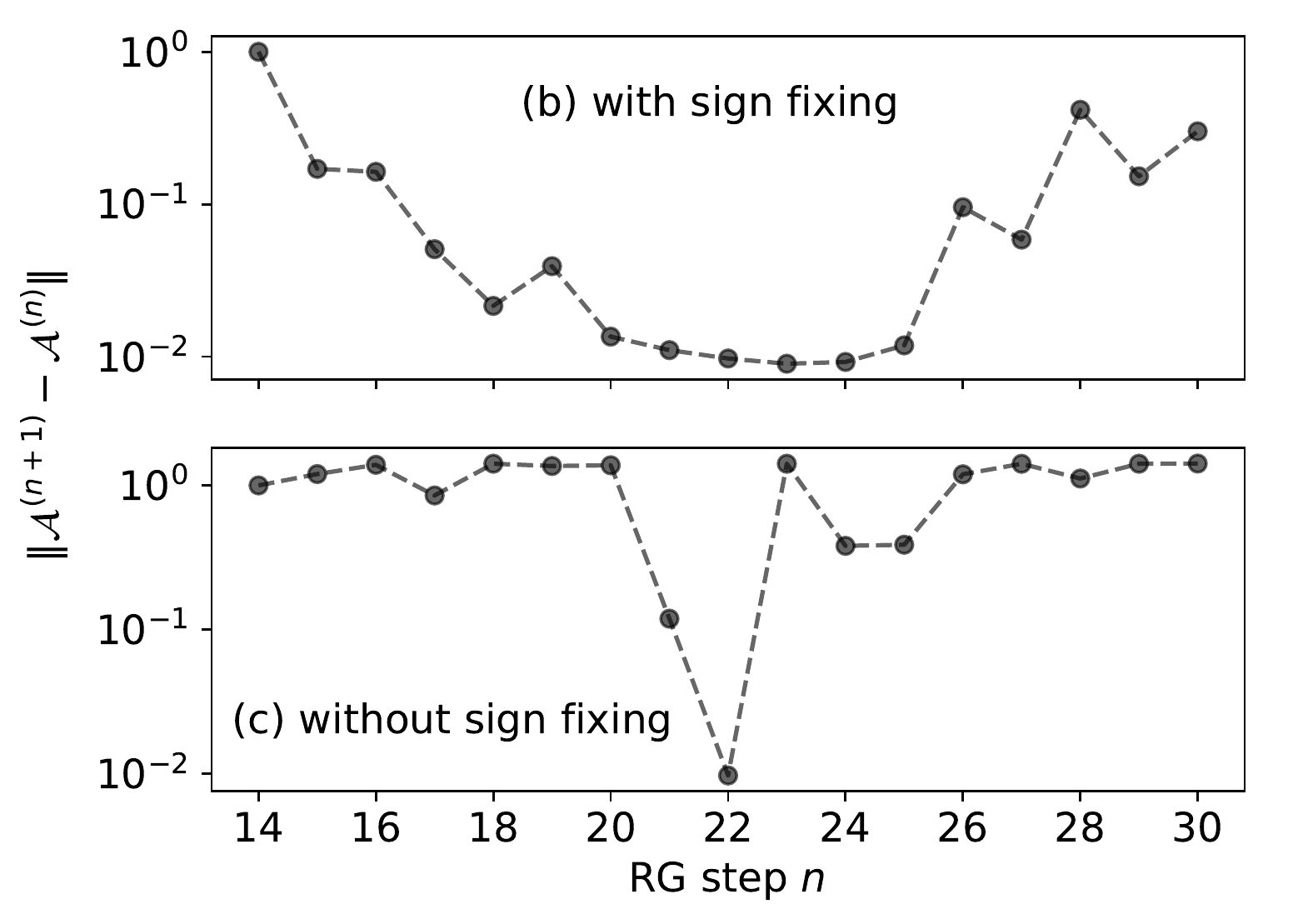}
    \caption{\label{fig:flowA}
        The RG flows of (a) singular values defined in Eq.~\eqref{def:Asvd} and (b) the difference between the normalized tensors, $\Vert \mathcal{A}^{(n+1)} - \mathcal{A}^{(n)} \Vert$ with sign fixing and (c) without, all at the estimated critical temperature $T_c^{[30]}$, generated by the proposed HOTRG-like scheme with $\chi = 30, \epsilon_{\text{gilt}} = 6\times 10^{-6}$.
    }
\end{figure}

To make sure that the plateau in the RG flow of $\Vert A^{(n)} \Vert$ gives a critical fixed-point tensor $(A^{[30]})^*_{\text{cr}}$ at the estimated critical temperature $T_c^{[30]}$, we plot the singular values $s^{(n)}$ of tensors $\mathcal{A}^{(n)}$ defined as
\begin{align}\label{def:Asvd}
    \includegraphics[scale=1.0,valign=c]{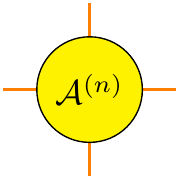}
    \svdeq
    \includegraphics[scale=1.0,valign=c]{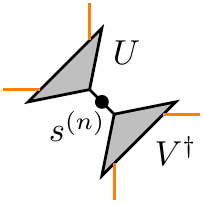}.
\end{align}
The RG flow of the singular values in Fig.~\ref{fig:flowA}(a) indicates that we indeed reach a non-trivial fixed-point tensor. 
The fixed-point tensor is manifestly fixed after adding the sign-fixing step, which can be confirmed by plotting the Frobenius norm of the difference between the normalized tensors at successive RG steps $\Vert \mathcal{A}^{(n+1)} - \mathcal{A}^{(n)}\Vert$, see Fig.~\ref{fig:flowA}(b). 
The norm of the difference starts to decay systematically at RG step $n = 14$, goes all the way down to the order $\sim 10^{-2}$ at $n = 23$ and then increases when the tensor begins to flow away from the critical fixed point.
By comparison, we show the RG flow of $\Vert \mathcal{A}^{(n+1)} - \mathcal{A}^{(n)}\Vert$ without sign fixing in Fig.~\ref{fig:flowA}(c); the sign ambiguities in Eq.~\eqref{eq:signAmbi} prevent us from achieving a manifestly-fixed-point tensor, except at RG step $n = 22$, where the tensor happens to have all signs correct by accident.
%%% Figure: the RG flows of singular values of tensors
% end Figure

\begin{figure}[tb]
    \includegraphics[width=\columnwidth]{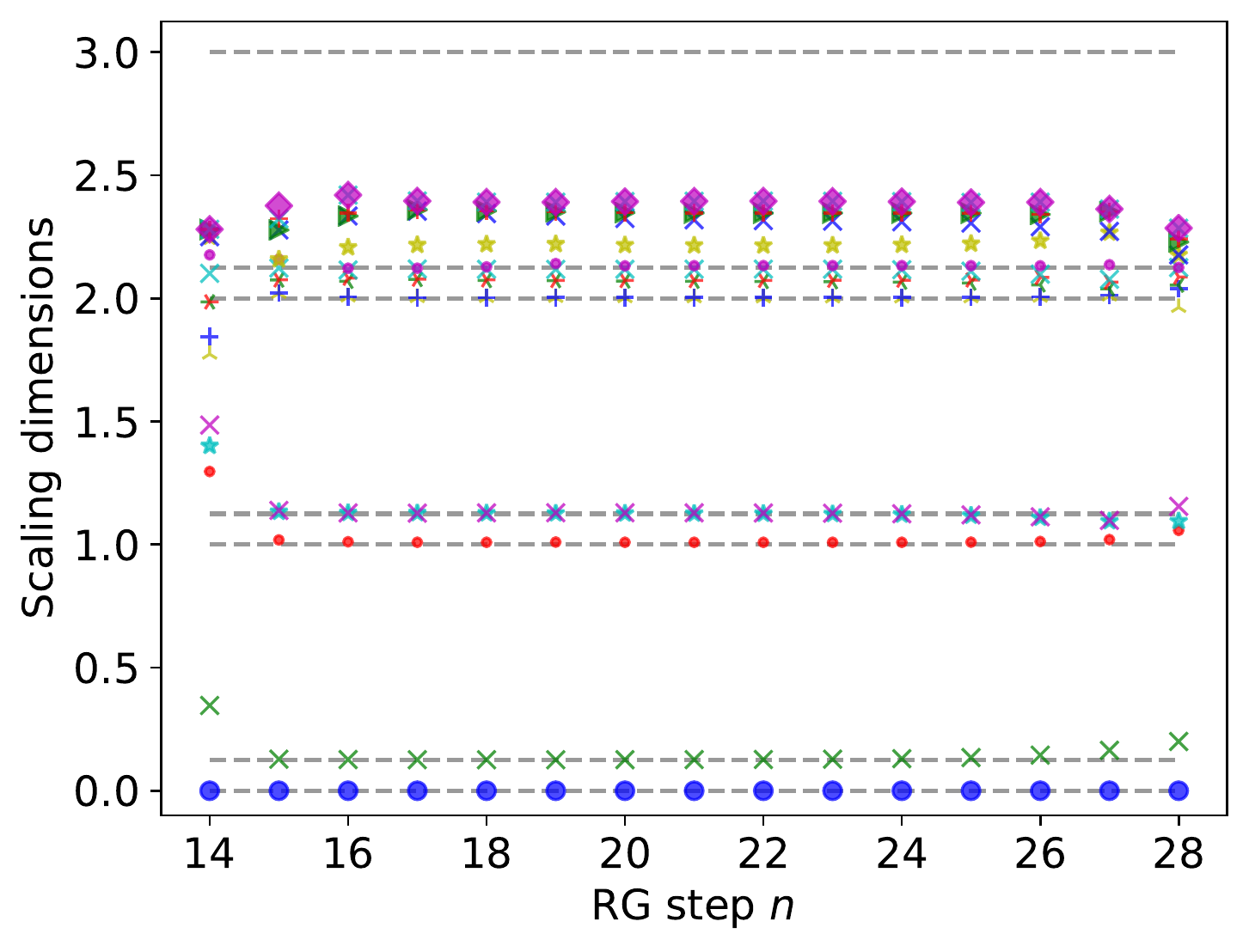}
    \caption{\label{fig:scDim}
        The scaling dimensions of the 2D Ising model from the canonical RG prescription using the proposed HOTRG-like scheme with $\chi = 30, \epsilon_{\text{gilt}} = 6\times 10^{-6}$.
    Dashed lines are the exact values.}
\end{figure}

We use the automatic differentiation implemented in JAX~\cite{jax2018github} to generate the linearized tensor RG equation $\mathcal{R}$ in Eq.~\eqref{eq:respMatGiltHOTRG} at RG steps $n = 14,15,\ldots, 28$, when the tensor is very close to the critical fixed-point tensor. 
The scaling dimensions are extracted from the eigenvalues of the matrix $\mathcal{R}$ according to Eq.~\eqref{eq:lambda2x}, where $b = 2, d = 2$. 
In Fig.~\ref{fig:scDim}, we show the first few scaling dimensions. 
The dashed lines are the exact values~\cite{DiFrancesco1997}. 
For $\chi = 30$, the RG prescription in tensor space gives correct scaling dimensions up to $2.125$. 
The results at RG step $n = 14 \text{ and } 28$ are unreliable since $\Vert \mathcal{A}^{(n+1)} - \mathcal{A}^{(n)}\Vert$ is of order $1$ (see Fig.~\ref{fig:flowA}(b)). 
The results for $n = 15,16,\ldots,27$ indicates that the scaling dimensions from the RG prescription in tensor space are reliable as long as the values of $\Vert \mathcal{A}^{(n+1)} - \mathcal{A}^{(n)}\Vert$ have order of or smaller than $10^{-1}$. 
%
%%% Figure: scaling dimensions
%%% end figure

In Table~\ref{table:scDim}, we show the scaling dimensions for all relevant and marginal operators at RG step $n = 22$ from the canonical RG prescription, compared with the results obtained by Gu and Wen's method~\cite{GuWen2009}, where two copies of the fixed-point tensor are used to construct the transfer matrix. 
Both methods have similar accuracy for scaling dimensions less than or equal to $1.125$.
The RG prescription in tensor space gives two out of total four scaling dimensions $2$ with three digits of accuracy, but the remaining two are overestimated and closer to $2.125$.
Gu and Wen's method estimates all of the four scaling dimensions $2$ correctly with two digits of accuracy.
%
%%% Table: scaling dimensions
\begin{table}[t]%[H] add [H] placement to break table across pages
\caption{The scaling dimensions for the relevant and marginal operators
    of the 2D Ising model at criticality from the canonical RG prescription 
    and from the transfer matrix method \`a la Gu and
    Wen~\cite{GuWen2009}, both using the proposed HOTRG-like scheme with $\chi = 30, \epsilon_{\text{gilt}} =
    6\times 10^{-6}$ at RG step $n = 22$.\label{table:scDim}} 
\begin{ruledtabular}
\begin{tabular}{ c c c c c c c c c }
Exact      & 0.125 & 1 & 1.125 & 1.125 & 2 & 2 & 2 & 2 \\
\hline
\thead{RG\\ pres.} & 0.127 & 1.009 & 1.125 & 1.128 & 2.002 &
2.004 & 2.068 & 2.073 \\
\thead{Trans.\\ mat.} & 0.125 & 1.002 & 1.128 & 1.128 & 2.014 &
2.014 & 2.016 & 2.016
\end{tabular}
\end{ruledtabular}
\end{table}
% end Table

We end this section with a few remarks on the above calculations. 
Firstly we impose the $\mathbb{Z}_2$ symmetry of the tensors~\cite{Singh2010SymTen, Singh2011U1Ten} when generating the RG flow in tensor space. 
There are three reasons. 
Only if the $\mathbb{Z}_2$ symmetry of the tensor is imposed will the low-temperature fixed-point tensor be stable under the RG.  
Otherwise, it will flow to the high-temperature fixed point eventually due to numerical errors, which will make the bisection search for the estimated critical temperature $T_c^{[\chi]}$ less convenient. 
The second merit of symmetric tensors is that half of the gauge redundancy can be automatically fixed (see Sec.~\ref{sec:gaugefix}), making the sign-fixing procedure easier. 
The third reason is to speed up the computations. 
However, we roll back to ordinary tensors when performing the RG prescription in tensor space, since the perturbations around the fixed-point tensor do not have to preserve $\mathbb{Z}_2$ symmetry (for example, the spin operator)\footnote{
    The perturbations like the spin operator are not $\mathbb{Z}_2$ invariant, but they are $\mathbb{Z}_2$ covariant, with nonzero charge. 
    It should be possible to utilize this numerically to extract scaling dimensions of operators with different charges separately.
    However, we did not proceed in this direction here.
}. 

The second remark is about the improvement of the accuracy as the bond dimension $\chi$ increases. 
There are two sources of approximation errors in the above computations. 
One comes from the truncations of the CDL tensors during the GILT that is necessary for producing the critical fixed point. 
This error is controlled by the hyper-parameter $\epsilon_{\text{gilt}}$. 
The other source is the leg squeezing step during the HOTRG to prevent the grow of the bond dimension.
This error can be reduced by increasing the bond dimension $\chi$. 
In general, for a given $\chi$, the $\epsilon_{\text{gilt}}$ should be as small as possible provided that the proposed HOTRG-like scheme can exhibit a critical fixed-point tensor. 
In practice, we tried $\chi = 10, 20, 30$, and $\epsilon_{\text{gilt}}$ goes down from $6\times 10^{-4}$ to $6\times10^{-5}$ and further to $6\times10^{-6}$. 
The estimated scaling dimensions converge to the exact results in this process. 

The third remark is about the overall multiplication constant in front of the fixed-point tensor. 
After reaching the critical fixed point, the RG from $n$-th step to $(n+1)$-th step is the map $\mathcal{A}^{*}\rightarrow c^{*} \mathcal{A}^{*}$, where $c^{*}$ is the magnitude of the coarser tensor. 
The shape of $\mathcal{A}^*$ is fixed but its magnitude is still changing under the RG transformation. 
It has been shown in Ref.~\cite{GuWen2009} that the fixed-point tensor with correct magnitude is simply given by $A^* = (c^*)^{-1/3} \mathcal{A}^*$, and we will have $A^*\rightarrow A^*$ under the RG transformation.
Our numerical results have confirmed this statement.

The final remark is that the problem of local correlations could be removed by other methods~\cite{GuWen2009,tnr,tnralgo,tnrplus,looptnr,harada2018,fet,tns,tensor-ring} other than GILT.
For example, the TNR~\cite{tnr,tnralgo} is known to be capable of exhibiting critical fixed-point tensors with its RG equation similar to that of the proposed HOTRG-like scheme in Eq.~\eqref{def:RGeqGiltHOTRG}, and there is a method to fix its gauge~\cite{tnralgo}. 
Considering the unprecedented accuracy of the TNR, the estimation of the scaling dimensions might be much better.
We develop the canonical RG prescription in tensor space using the HOTRG-like scheme in this paper in order to prepare for the further applications to 3D systems.

\section{Summary and discussions\label{conclusion}}
In this paper, we show how to perform the canonical RG prescription in tensor space.
The general procedure is summarized as follows: reach a fixed-point tensor using a tensor RG equation free of the problem of local correlations, fix the gauge redundancy to make the fixed-point tensor manifestly fixed, linearize the RG equation around this fixed-point tensor and finally calculate the scaling dimensions from the eigenvalues of this linearized tensor RG equation.
In practice, we propose an HOTRG-like scheme to carry out this canonical RG prescription in tensor space.
For the estimates of the scaling dimensions, we had not expected the present scheme would yield better accuracy than the conventional way \`a la Gu and Wen, and indeed it turned out not to be the case for the 2D classical Ising model.
However, the important fact is that the present scheme works at least equally well, and it potentially has a broader range of applications.

The success of the canonical RG prescription in tensor space offers a crucial missing piece of puzzle for understanding TRG-type techniques as real-space RG transformations.
The realization of the RG prescription based on the proposed HOTRG-like scheme extends the old Migdal-Kadanoff idea, and is systematically improvable.
The distinctive feature of the proposed method, compared with the two existing tensor-RG-based ones for extracting scaling dimensions~\cite{GuWen2009,EvenblyDilatationOp}, is that it is from pure RG perspective and does not rely on any CFT arguments explicitly, making the method more promising in 3D.
In our future work, we will generalize the HOTRG-like scheme and apply the canonical tensor RG prescription to 3D systems, where there are few practical tensor-network-based methods\footnote{
    From the perspective of a real-space RG transformation for quantum systems~\cite{ER2007}, the scale-invariant multiscale entanglement renormalization ansatz (MERA)~\cite{MERA} can be used to build a scaling superoperator~\cite{MERAsupop}. 
    The scaling dimensions are obtained from the eigenvalues of the scaling superoperator~\cite{supop2scaleD}. 
However, the computation costs of the MERA for 2+1D quantum systems grow as $O(\chi^{16})$~\cite{MERA2p1}, much higher than $O(\chi^{11})$ for the 3D HOTRG\@.
} 
to extract scaling dimensions efficiently.

\begin{acknowledgments}
We thank Satoshi Morita, Shumpei Iino, Takuhiro Ogino, Yuan Yao and Takeo Kato for fruitful discussions and insightful suggestions, and Glen Evenbly and Guifre Vidal for explanations regarding the TNR and other tensor network methods. 
We also thank Markus Hauru for clarifying the implementation of the GILT, and are very grateful to Antoine Tilloy and an anonymous referee for useful suggestions about this manuscript.
X.L.\ and R.G.X.\ are grateful to the support of the Global Science Graduate Course (GSGC) program of the University of Tokyo. 
This work is financially supported by MEXT Grant-in-Aid for Scientific Research (B) (19H01809).
The numerical computations were performed on computers at the Supercomputer Center, the Institute for Solid State Physics (ISSP), the University of Tokyo.

\end{acknowledgments}

% appendices.
\appendix
\section{GILTs designed for the HOTRG\label{append:gilthotrg}}
%
%% Gilt applied on plaquettes
\begin{figure*}[!t]
\includegraphics[scale=1.0,valign=c]{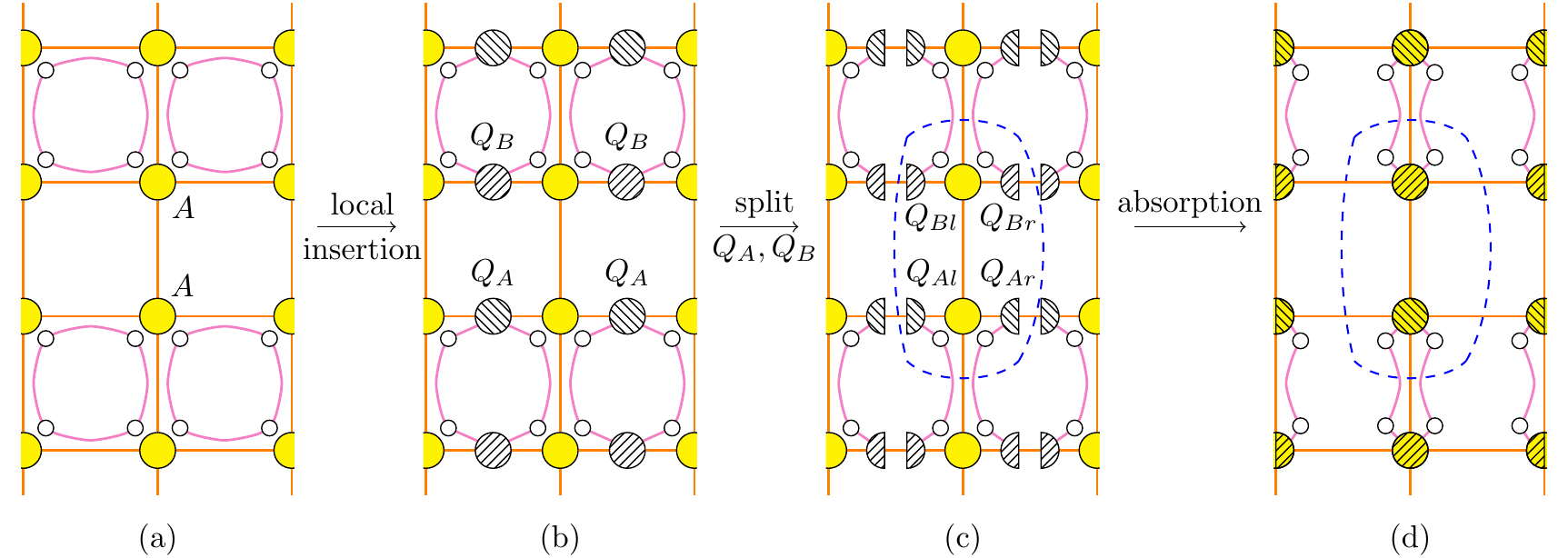}
\caption{\label{fig:gilt4hotrg}
    The plaquettes and bonds where the GILT is applied to for the subsequent HOTRG coarse graining.
    (a) Two copies of $A$ in the center will be coarse grained vertically. 
    The problematic loops of local correlations are drawn explicitly in the plaquettes to make the demonstration clearer.
    They are unknown inner structure of the main tensor $A$.
    (b) Copies of low-rank matrices $Q_A,Q_B$, determined by the GILT~\cite{gilts}, are inserted into the bonds to catch the legs of the loops.
    (c) $Q_A,Q_B$ are split using singular value decomposition. The GILT ensures the legs of the loops do not leak out.
    (d) The pieces of $Q_A,Q_B$ matrices are absorbed into the copies of tensor $A$. The subsequent HOTRG will be applied on the patch of tensors in the dashed circle.
}
\end{figure*}

In this appendix, we first briefly introduce how the low rank matrix is determined in the GILT, and then move on to explain why the HOTRG has difficulty in filtering out the local correlations and how the GILT comes to help.

The low-rank matrix $Q$ in Fig.~\ref{fig:gilt} is determined by examining the environment $E$ of the bond and performing the singular value decomposition,
\begin{align}\label{eq:bondEnvSVD}
    E \equiv 
    \includegraphics[scale=1.0,valign=c]{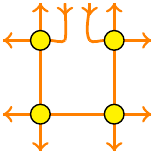}
    \svdeq 
    \includegraphics[scale=1.0,valign=c]{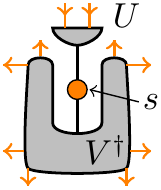}\text{ },
\end{align}
where we refrain from drawing the unknown $C$ matrices in the plaquette.
The environment $E$ of the bond should be thought of as a linear map from the vector space of all the legs with ingoing arrows to that of all the legs with outgoing arrows. 
We can use the tensor $U$ and the diagonal matrix $s$ in Eq.~\eqref{eq:bondEnvSVD} to construct the low-rank matrix $Q$. 
To this end, we first define a vector $t$ by contracting two ingoing legs of the tensor $U$,
\begin{align}\label{def:tfromU}
    \includegraphics[scale=1.0,valign=c]{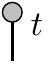}
    \equiv 
    \includegraphics[scale=1.0,valign=c]{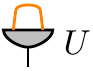}.
\end{align}
Then, we perform a soft truncation of the vector $t$ according to
\begin{align}\label{eq:tpfromt}
    t'_i = t_i \frac{s_i^2}{s_i^2 + \epsilon_{\text{gilt}}^2},
\end{align}
where $s_i$ are the singular values and $\epsilon_{\text{gilt}}$ is the hyper-parameter of the GILT.
Equation~\eqref{eq:tpfromt} says that the components of the vector $t$ will be set to very small values if the corresponding singular values $s_i$ are much smaller than $\epsilon_{\text{gilt}}$. 
The justification for the truncation in Eq.~\eqref{eq:tpfromt} can be found in Ref.~\cite{gilts}. 
The low-rank matrix $Q$ is constructed from the tensor $U^{\dagger}$ and the truncated vector $t'$ as
\begin{align}\label{def:QfromUtp}
    \includegraphics[scale=1.0,valign=c]{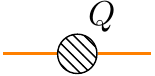}
    \equiv 
    \includegraphics[scale=1.0,valign=c]{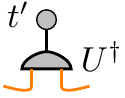}\text{ }.
\end{align}
It is proved in Ref.~\cite{gilts} that the matrix $Q$ determined in this
way is able to filter out the loop of four $C$ matrices shown in Fig.~\ref{fig:gilt}. 

Next, we demonstrate how  to choose the plaquettes and where to insert the low-rank matrices to filter out the unwanted local correlations for the HOTRG.
It is shown in Ref.~\cite{hotrgfixpoint} that the HOTRG in the vertical direction transforms the $A^{\text{CDL}}$ in Fig.~\ref{fig:rgschem}(b) in the following way,
\begin{align}\label{eq:cdlHOTRG}
    \includegraphics[scale=0.8,valign=c]{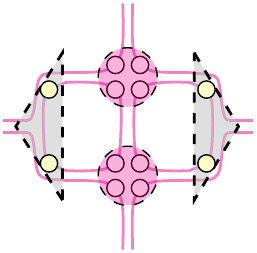}
    \propto 
    \includegraphics[scale=1.0,valign=c]{singleCDL.pdf},
\end{align}
which means that although the HOTRG can detect and project out four inner $C$ matrices, it can do nothing about the four outer $C$ matrices.
Therefore, the GILT should be applied to filter out these four outer $C$ matrices before the HOTRG coarse graining. 
To this end, we apply the GILT to the plaquettes where the loops of local correlations are drawn explicitly in Fig.~\ref{fig:gilt4hotrg} and insert two low-rank matrices $Q_A,Q_B$ into the upper and lower bonds for each plaquette. 
The legs of the unwanted $C$ matrices will be truncated after the splitting of $Q_A, Q_B$. 
Finally, we apply the ordinary HOTRG in the vertical direction to the local patch of tensors in the dashed circle in Fig.~\ref{fig:gilt4hotrg} to get the coarser tensor $A'$
\begin{align}\label{def:ApycontrGilt}
    \includegraphics[scale=1.0,valign=c]{Ap-ycontr.pdf}
    \equiv
    \includegraphics[scale=0.8,valign=c]{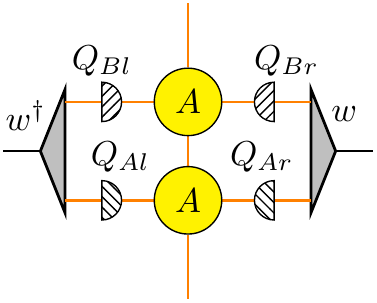}.
\end{align}
In this way, we can remove all horizontal legs of $C$ matrices: a half of them by the GILT and the other half by contraction in the HOTRG\@.
We repeat the similar GILT and the HOTRG on $A'$ in the horizontal direction.
The coarse-graining steps in two directions together define the tensor RG equation of the HOTRG-like scheme in Eq.~\eqref{def:RGeqGiltHOTRG}.

\section{Proof regarding gauge fixing\label{append:proofgaugefix}}
To see why the gauge fixing procedure in Eqs.~\eqref{def:Nx} to~\eqref{def:gaugefixHori} defines a preferred set of basis, let us examine how the tensor $\tilde{A}$ in Eq.~\eqref{def:gaugeTrans} transforms under this gauge fixing procedure.
The contraction of two vertical legs of $\tilde{A}$ annihilates $S_y$ and $S_y^{-1}$ on the right hand side of Eq.~\eqref{def:gaugeTransPic}; the resultant $\tilde{N}_x$ is related to $N_x$ through
\begin{align}\label{eq:simTransNx}
    \includegraphics[scale=0.8,valign=c]{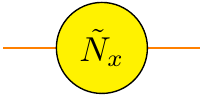}
    = 
     \includegraphics[scale=0.8,valign=c]{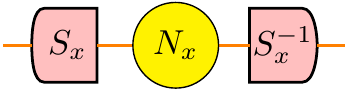}.
\end{align}
Provided that there is no degeneracy in the eigenvalue spectrum of $N_x$, the matrix $\tilde{W}_x$ coming from eigenvalue decomposition of the matrix $\tilde{N}_x$ is related to $W_x$ in Eq.~\eqref{eq:eigdcpNx} through
\begin{align}\label{eq:Wxtransf}
    \includegraphics[scale=0.8,valign=c]{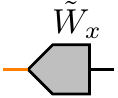}
    = 
    \includegraphics[scale=0.8,valign=c]{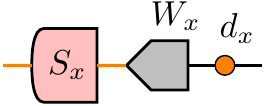},
\end{align}
where $d_x$ is a diagonal matrix coming from phase ambiguities of eigenvectors, with its diagonal entries to be phases for general complex matrices. 
For a real symmetric $N_x$, the diagonal entries of $d_x$ are $\pm 1$.
After the horizontal gauge fixing, the tensor $\tilde{A}$ becomes
\begin{align}\label{eq:AtildegaugefixHori}
    \includegraphics[scale=0.8,valign=c]{Atilde.pdf}
    \xrightarrow[\text{gauge fixing}]{\text{horizontal}} 
    \includegraphics[scale=0.8,valign=c]{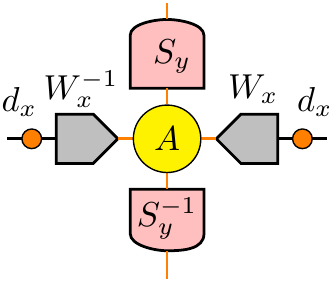}.
\end{align}
Compare Eq.~\eqref{def:gaugefixHori} with Eq.~\eqref{eq:AtildegaugefixHori}, we see that the gauge redundancies in two horizontal legs are fixed except the phase ambiguities.
For 2D classical statistical models with spatial reflection symmetries, for example, the 2D Ising model, the real matrix $N_x$ can be made symmetric, so the phase ambiguities become sign ambiguities.

Finally, let us prove the property of the tensor RG equation of the HOTRG-like scheme in Eq.~\eqref{eq:signAmbi}.
We focus on real tensors (the generalization to complex tensors is straightforward) and the equivalence relation defined by the gauge transformation,
\begin{align}\label{eq:simorth}
    \includegraphics[scale=0.8,valign=c]{Atilde.pdf}
    = 
    \includegraphics[scale=0.8,valign=c]{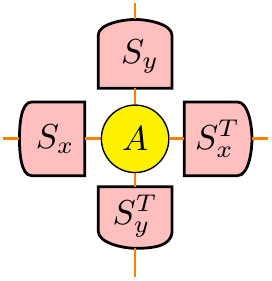},
\end{align}
where $S_x,S_y$ are orthogonal matrices.
It is sufficient to consider such orthogonal changes of gauge if we restrict to the representations of the equivalence class $[A]$ with spatial reflection symmetries~\cite{tnralgo},
\begin{subequations}\label{def:relsym}
    \begin{align}    A_{kjil}  = \sum_{j'l'} (O_y)_{j j'} (O_y)_{l l'} A_{ij'kl'} 
    \end{align}
and
    \begin{align}        
        A_{ilkj} = \sum_{i'k'} (O_x)_{i i'} (O_x)_{k k'} A_{i'jk'l},
    \end{align} 
\end{subequations}
where $O_x, O_y$ are orthogonal matrices, also with $O_x^2 = O_y^2 = \mathbb{1}$, and the legs' order convention is as per Eq.~\eqref{def:tensorA}.
It can be shown that, if we start with a tensor with the reflection symmetries, the tensor RG equation of the HOTRG-like scheme in Eq.~\eqref{def:RGeqGiltHOTRG} will preserve the reflection symmetries\footnote{This is because the pieces of low-rank matrices and the isometric tensors in the RG equation of the HOTRG-like scheme in Eq.~\eqref{def:RGeqGiltHOTRG} will inherit the reflection symmetries of the input tensor $A$.
}
and will rotate the tensor into the set of basis where $O_x, O_y$ become diagonal, with their diagonal entries $\pm 1$.

It suffices to discuss the first half of the coarse graining defined in Eq.~\eqref{def:ApycontrGilt}. 
We want to show that if $\tilde{A}$ is fed into the right hand side of Eq.~\eqref{def:ApycontrGilt}, the $\tilde{A}'$ we obtain on the left hand side is related with the original $A'$ by 
\begin{align}\label{eq:Ap2Aptilde}
    \includegraphics[scale=0.8,valign=c]{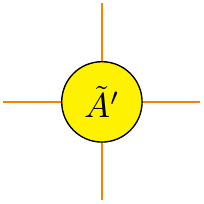}
    = 
    \includegraphics[scale=0.8,valign=c]{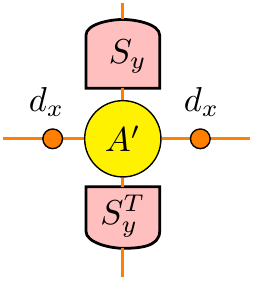},
\end{align}
where $d_x$ is a diagonal matrix with diagonal entries $\pm 1$.
Equation~\eqref{eq:Ap2Aptilde} means that the gauge redundancy in the horizontal legs will be fixed with only sign ambiguities left during the first half of the coarse graining in the vertical direction.
It follows immediately that the full tensor RG equation in Eq.~\eqref{def:RGeqGiltHOTRG} will give Eq.~\eqref{eq:signAmbi}.

Let us first figure out the correct $\tilde{Q}_A,\tilde{Q}_B$ matrices in Fig.~\ref{fig:gilt4hotrg}. 
The environment in Eq.~\eqref{eq:bondEnvSVD} is multiplied by several orthogonal matrices, which will not change the singular values, so $\tilde{s}_i=s_i$. 
It is easy to check that the tensor $U$ in the singular value decomposition becomes (the sign ambiguities coming from the singular value decomposition does not matter here)
\begin{align}\label{eq:Utilde}
    \includegraphics[scale=0.8,valign=c]{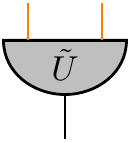}
    = 
    \includegraphics[scale=0.8,valign=c]{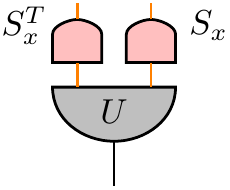}.
\end{align}
The vector $\tilde{t}$ is thus the same as the original $t$ by its definition in Eq.~\eqref{def:tfromU}, which further gives $\tilde{t}'_i= t'_i$ since the tilde version of the right hand side of Eq.~\eqref{eq:tpfromt} is the same as the original version. 
Finally, equation~\eqref{def:QfromUtp} gives
\begin{align}\label{eq:QAtilde}
    \includegraphics[scale=1.0,valign=c]{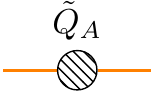}
    \texteq{\eqref{def:QfromUtp}}
    \includegraphics[scale=1.0,valign=c]{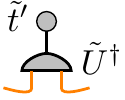}
    &= 
    \includegraphics[scale=0.8,valign=c]{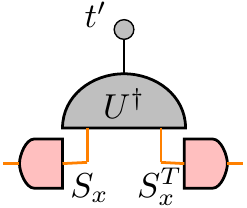}
    \nonumber\\ 
    &= 
    \includegraphics[scale=1.0,valign=c]{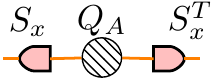}.
\end{align}
Equation~\eqref{eq:QAtilde} means that the low rank matrix $Q_A$ transforms in a nice way when we perform a gauge transformation defined in Eq.~\eqref{eq:simorth}. 
If the singular values of $Q_A$ \textit{do not have degeneracy}, after splitting of $Q_A$, we have $Q_{Ar},Q_{Al}$ transform like (the sign ambiguities coming from singular value decomposition of $Q_A$ and $\tilde{Q}_A$ would kick in and contribute to $d_x$ in Eq.~\eqref{eq:Ap2Aptilde}, but they are not drawn explicitly in the equation below)
\begin{align}\label{eq:QpieceTrans}
    \includegraphics[scale=1.0,valign=c]{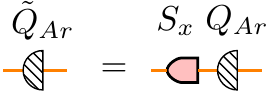}
    \text{ and } 
    \includegraphics[scale=1.0,valign=c]{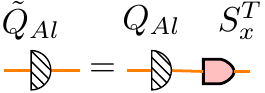}.
\end{align}
The $S_x,S_x^T$ matrices that $Q_{Ar},Q_{Al}$ pick up will cancel those acting on the $A$ tensor when $\tilde{Q}_{Ar},\tilde{Q}_{Al}$ are contracted with the $\tilde{A}$ tensor in Eq.~\eqref{eq:simorth}.
The same argument works for $Q_B$. 
Equation~\eqref{eq:QpieceTrans} indicates that all the $S_x,S_x^T$ matrices acting on the four horizontal legs of the local patch in Eq.~\eqref{def:ApycontrGilt} will be canceled by the low-rank matrices used in the GILT process. 
The above analysis shows that during the GILT process for the vertical coarse graining, the gauge in the horizontal legs will be fixed with only sign ambiguities left, since the GILT favors the basis chosen by the singular value decompositions of $Q_A, Q_B$.

However, there is one more twist. In practice, we observe that the low-rank matrices are projection operators, which are highly degenerated. 
As a result, the gauge redundancy in the degenerate subspace will leak out, which will be seen by the subsequent HOTRG process. 
Luckily, the HOTRG has a similar feature as the GILT process.
It favors the basis where the positive semi-definite matrix $M M^{\dagger}$ (see the definition of matrix $M$ in Eq.~\eqref{def:M-AA}) is diagonal (the sign ambiguities coming from eigenvalue decomposition of $M M^{\dagger}$ would similarly kick in here and contribute to $d_x$ in Eq.~\eqref{eq:Ap2Aptilde}). 
It is straightforward to see that the isometry $w$ will pick up the suitable $S_x,S_x^T$ matrices to cancel out the gauge transformation leaking out from the GILT process.
There are still concerns about whether degeneracy occurs in eigenvalues of $M M^{\dagger}$. 
Our result in Fig.~\ref{fig:flowA}(b) shows, a posteriori, that the potential degeneracy does not cause any problem for the 2D Ising model at criticality.

\section{Source Code\label{append:sc}}
The source code of this paper can be found at \href{https://github.com/brucelyu/tensorRGflow}{github.com/brucelyu/tensorRGflow}.
It can be used to reproduce all the results in Sec.~\ref{benchmark:2DIsing} for the 2D classical Ising model.

\bibliography{tensorRGflow}

\end{document}